\documentclass[10pt, conference, letterpaper]{IEEEtran}
\usepackage{graphicx}
\usepackage{xspace}
\usepackage{color}
\usepackage{paralist}
\usepackage{wrapfig}
\usepackage{multirow}
\usepackage{listings}
\usepackage{makecell}
\usepackage{boxedminipage}
\usepackage{booktabs}
\usepackage{subcaption}
\usepackage{caption}
\usepackage{diagbox}
\usepackage{adjustbox}
\usepackage{url}
\usepackage{tabularx}
\usepackage{breakurl} 
\usepackage{pifont}
\usepackage{textcomp}
\usepackage{balance} 
\usepackage{upquote}
\usepackage[T1]{fontenc}
\usepackage{algorithm}
\usepackage{algpseudocode}

\usepackage{pifont}
\usepackage{balance}
\usepackage[htt]{hyphenat}
\usepackage[table,xcdraw]{xcolor} % For color support
\usepackage[colorlinks=true,linkcolor=blue,citecolor=red,urlcolor=blue]{hyperref} % Customize link colors

\usepackage[inline]{enumitem}

\makeatletter

\newcommand{\refappendix}[1]{\hyperref[#1]{Appendix~\ref*{#1}}}

\begin{document}
\newcommand{\bheading}[1]{{\vspace{2pt}\noindent{\textbf{#1}}}}
\newcommand{\iheading}[1]{{\vspace{2pt}\noindent{\textit{#1}}}} 

%terminology and formatting
\newcounter{note}[section]
\renewcommand{\thenote}{\thesection.\arabic{note}}
\newcommand{\yz}[1]{\refstepcounter{note}{\bf\textcolor{green}{$\ll$YZ~\thenote: {\sf #1}$\gg$}}}
\newcommand{\mz}[1]{\refstepcounter{note}{\bf\textcolor{red}{$\ll$MZ~\thenote: {\sf #1}$\gg$}}}
\newcommand{\xz}[1]{\refstepcounter{note}{\bf\textcolor{blue}{$\ll$XZ~\thenote: {\sf #1}$\gg$}}}
\newcommand{\jn}[1]{\refstepcounter{note}{\bf\textcolor{violet}{$\ll$JN~\thenote: {\sf #1}$\gg$}}}
\newcommand{\xl}[1]{\refstepcounter{note}{\bf\textcolor{orange}{$\ll$XL~\thenote: {\sf #1}$\gg$}}}
\newcommand{\ZQ}[1]{\refstepcounter{note}{\bf\textcolor{red}{$\ll$ZQ~\thenote: {\sf #1}$\gg$}}}

\newcommand\st[1]{\ding{#1}}

\renewcommand{\figurename}{Fig.}

\newcommand{\etal}{\emph{et al.}\xspace}
\newcommand{\etc}{\emph{etc}\xspace}
\newcommand{\ie}{\emph{i.e.}\xspace}
\newcommand{\eg}{\emph{e.g.}\xspace}

\newcommand{\figurewidth}{\columnwidth}
\newcommand{\secref}[1]{\mbox{\S\ref{#1}}\xspace}
\newcommand{\secrefs}[2]{\mbox{Sec.~\ref{#1}--\ref{#2}}\xspace}
\newcommand{\figref}[1]{\mbox{Fig.~\ref{#1}}}
\newcommand{\tabref}[1]{\mbox{Table~\ref{#1}}}
\newcommand{\appref}[1]{\mbox{Appendix~\ref{#1}}}
\newcommand{\ignore}[1]{}

% units
\newcommand{\gbytes}{\ensuremath{\mathrm{GB}}\xspace}
\newcommand{\mbytes}{\ensuremath{\mathrm{MB}}\xspace}
\newcommand{\kbytes}{\ensuremath{\mathrm{KB}}\xspace}
\newcommand{\bytes}{\ensuremath{\mathrm{B}}\xspace}
\newcommand{\hertz}{\ensuremath{\mathrm{Hz}}\xspace}
\newcommand{\ghertz}{\ensuremath{\mathrm{GHz}}\xspace}
\newcommand{\msecs}{\ensuremath{\mathrm{ms}}\xspace}
\newcommand{\usecs}{\ensuremath{\mathrm{\mu{}s}}\xspace}
\newcommand{\nsecs}{\ensuremath{\mathrm{ns}}\xspace}
\newcommand{\secs}{\ensuremath{\mathrm{s}}\xspace}
\newcommand{\gbits}{\ensuremath{\mathrm{Gb}}\xspace}

\newcounter{packednmbr}
\newenvironment{packedenumerate}{
\begin{list}{\thepackednmbr)}{\usecounter{packednmbr}
\setlength{\itemsep}{0pt}
\addtolength{\labelwidth}{4pt}
\setlength{\leftmargin}{12pt}
\setlength{\listparindent}{\parindent}
\setlength{\parsep}{3pt}
\setlength{\topsep}{3pt}}}{\end{list}}

\newenvironment{packeditemize}{
\begin{list}{$\bullet$}{
\setlength{\labelwidth}{8pt}
\setlength{\itemsep}{0pt}
\setlength{\leftmargin}{\labelwidth}
\addtolength{\leftmargin}{\labelsep}
\setlength{\parindent}{0pt}
\setlength{\listparindent}{\parindent}
\setlength{\parsep}{0pt}
\setlength{\topsep}{3pt}}}{\end{list}}

% paper specific macros
\newcommand{\sysname}{\texttt{SysName}\xspace}

\newcommand{\lf}{\textit{lockout factors}\xspace}

\newcommand{\threshold}{\ensuremath{\mathit{thrsh}}\xspace}

\newcommand{\cmark}{\ding{51}}%
\newcommand{\xmark}{\ding{55}}%

% general concept
\newcommand{\tx}{transaction\xspace}
\newcommand{\Tx}{Transaction\xspace}
\newcommand{\txs}{transactions\xspace}
\newcommand{\Txs}{Transactions\xspace}
\newcommand{\mev}{MEV\xspace}
\newcommand{\pri}{private\xspace}
\newcommand{\Pri}{Private\xspace}
\newcommand{\nl}{normal\xspace}
\newcommand{\Nl}{Normal\xspace}
\newcommand{\Fb}{Flashbots\xspace}

\newcommand{\nalgref}[1]{\mbox{Alg.~\ref{#1}}}
\newcommand{\callindex}{\textsf{\small CallIndex}\xspace}
\newcommand{\callstack}{\textsf{\small CallStack}\xspace}

% %% lessons
% \newcounter{lessoncount}
% \newcommand{\takeaway}[4]{
% \refstepcounter{lessoncount}
% \vspace{4pt}
% % \setlength\fboxrule{0.8pt}
% \noindent\fbox{\parbox{0.96\linewidth}{
%     Observation~\thelessoncount: \IfEndWith{#1}{.}{#1}
% }}}

\newcounter{lessoncount}

\newcommand{\lesson}[1]{
\refstepcounter{lessoncount}
\vspace{2pt}
\setlength\fboxrule{0.8pt}
\noindent\fbox{%
\parbox{0.98\linewidth}{%
   \textbf{Finding~\thelessoncount:} {#1}
}}
\vspace{2pt}
}

\newcommand{\observation}[1]{
\vspace{4pt}
\setlength\fboxrule{0.8pt}
\noindent\fbox{%
\parbox{0.96\linewidth}{%
    {#1}
}}
\vspace{6pt}
}

\newcommand{\swap}{\textit{Swap}\xspace}
\newcommand{\swaps}{\textit{Swaps}\xspace}

\title{Demystifying Private Transactions and Their Impact in PoW and PoS Ethereum}

\author{
\IEEEauthorblockN{Xingyu Lyu}
\IEEEauthorblockA{
% Computer Science and Engineering\\
\textit{University of Massachusetts, Lowell}\\
Lowell, Massachusetts, USA\\
xingyu\_lyu@uml.edu}
\and
\IEEEauthorblockN{Mengya Zhang}
\IEEEauthorblockA{
% Computer Science and Engineering\\
\textit{The Ohio State University}\\
Columbus, Ohio, USA \\
zhang.9407@osu.edu}
\and
\IEEEauthorblockN{Jianyu Niu}
\IEEEauthorblockA{
\textit{Southern University of Science and Technology}\\
Shenzhen, CHINA\\
niujy@sustech.edu.cn}
\and
\IEEEauthorblockN{Xiaokuan Zhang}
\IEEEauthorblockA{
\textit{George Mason University}\\
Fairfax, VA, USA \\
xiaokuan@gmu.edu}
\and
\IEEEauthorblockN{Yinqian Zhang}
\IEEEauthorblockA{
\textit{Southern University of Science and Technology}\\
% \textit{Research Institute of Trustworthy Autonomous Systems}\\
Shenzhen, CHINA\\
yinqianz@acm.org}
\and
\IEEEauthorblockN{Zhiqiang Lin}
\IEEEauthorblockA{
\textit{The Ohio State University}\\
Columbus, Ohio, USA \\
zlin@cse.ohio-state.edu}
}

\maketitle
\pagestyle{plain}

\begin{abstract}
In Ethereum, private transactions, a specialized transaction type employed to evade public Peer-to-Peer (P2P) network broadcasting, remain largely unexplored, particularly in the context of the transition from Proof-of-Work (PoW) to Proof-of-Stake (PoS) consensus mechanisms. To address this gap, we investigate the transaction characteristics, (un)intended usages, and monetary impacts by analyzing large-scale datasets comprising 14,810,392 private transactions within a 15.5-month PoW dataset and 30,062,232 private transactions within a 15.5-month PoS dataset. While originally designed for security purposes, we find that private transactions predominantly serve three distinct functions in both PoW and PoS Ethereum: extracting Maximum Extractable Value (MEV), facilitating monetary transfers to distribute mining rewards, and interacting with popular Decentralized Finance (DeFi) applications. Furthermore, we find that private transactions are utilized in DeFi attacks to circumvent surveillance by white hat monitors, with an increased prevalence observed in PoS Ethereum compared to PoW Ethereum. Additionally, in PoS Ethereum, there is a subtle uptick in the role of private transactions for MEV extraction. This shift could be attributed to the decrease in transaction costs. However, this reduction in transaction cost and the cancellation of block rewards result in a significant decrease in mining profits for block creators.
\end{abstract}


\section{Introduction}
\label{sec:intro}
Recent years have witnessed an explosive growth of Decentralized Finance (DeFi) applications in Ethereum, with a Total Value Locked (TVL) of around \$49.74 billion as of April 2024~\cite{defilama}. However, alongside this growth, there has been a noticeable rise in security breaches. Particularly, attackers have targeted the public mempool, where transactions await confirmation, to carry out frontrunning attacks. In such schemes, attackers monitor pending transactions in the mempool and strategically execute their transactions beforehand, aiming to secure profits at the expense of the original transaction initiators.

To mitigate these security vulnerabilities, particularly the threats posed by pending pool attacks, Ethereum pioneered the implementation of \textbf{\pri \txs} in its Proof-of-Work (PoW) period. This innovative approach allows for transactions to be sent directly to miners, effectively sidestepping the public mempool. By doing so, private transactions remain confidential, confined to the miner's exclusive domain, and thus insulated from the risks associated with public visibility and the consequent attacks.

Regrettably, private transactions have not been limited to protective use cases but have been widely adopted by malicious actors. These private transactions provide attackers with a means to evade being frontrun by transactions initiated by benevolent actors (e.g., whitehats). Additionally, private transactions have found utility in the extraction of Maximum Extractable Value (MEV)~\cite{mev}, which refers to the profits attainable through the manipulation of transaction order, inclusion, or censorship within a block. One prominent form of MEV is the sandwich attack, which involves both frontrunning and backrunning a victim's transaction. Given the intensifying competition for MEV extraction, private transactions are frequently leveraged to gain an advantage. For instance, MEV seekers bundle frontrunning and backrunning transactions with a victim's transaction and transmit them directly to miners through service providers like Flashbots~\cite{flashbot}. 

To guarantee the functionalities of \pri \txs, there comes an associated cost. Typically, every transaction carries a fee for its mining, including tips for miners. In addition to these standard fees, private transactions have been observed to involve direct payments to miners as incentives for prioritizing these transactions. For instance, consider an illustrative private transaction \texttt{0xc2969993} that sent 720.22 \textit{ETH} to its miner on June 13, 2022, with a value of approximately 0.72 million dollars at that time. The higher fee a private transaction has, the higher the likelihood of it being mined in the desired position, such as before a victim transaction, as miners are economically motivated to do so.

Notably, Ethereum underwent a significant switch on September 15, 2022, transitioning from a PoW to a Proof-of-Stake (PoS) consensus mechanism. While private transactions still rely on service providers, this shift has introduced substantial changes in how private transactions are managed. In the PoS Ethereum, a proposer-builder separation (PBS) approach has become prevalent, dividing miner roles into block builders and validators. This shift has prompted private transaction service providers to adapt their methods for handling private transactions. Specifically, providers bundled and forwarded private transactions to miners in PoW Ethereum, whereas in PoS Ethereum, they participate in building newly mined blocks and submit the best block to validators for proposal (see~\secref{sec:bg:pri} for details).

\begin{table*}[t]
    \begin{center}
    \resizebox{1\linewidth}{!}{
    \begin{tabular}{ cccccccccccccccccc } 
    \hline
    \toprule
      & \multicolumn{5}{c}{Transaction Dynamics} & & \multicolumn{3}{c}{Transaction Usages} & & \multicolumn{6}{c}{Economic Impacts}\\
      \cline{2-6} \cline{8-10} \cline{12-17}
      \multirow{2}{*}{} & \multirow{2}{*}{Count (m/m)} & \multicolumn{2}{c}{Distribution} & \multirow{2}{*}{Position} & \multirow{2}{*}{Fail (0 tip)} & \multirow{2}{*}{} & \multirow{2}{*}{MEV Bot} & \multirow{2}{*}{Top 3 Used Token Pair} & \multirow{2}{*}{Top 3 DeFi} & \multirow{2}{*}{} & \multicolumn{2}{c}{Transaction Cost (\textit{ETH})} & \multirow{2}{*}{} & \multicolumn{3}{c}{Mining Profits (\textit{ETH})} \\ 
      \cline{3-4} \cline{12-13} \cline{15-17}
      & & <= 10 & <= 20 & & & & & & & & Transaction Fee & Direct Payment & & Type I & Type II & Type III\\
      \midrule
    PoW & 0.95 (2.45\%) & 88.9\% & 96.3\% & 16 (186) & 2,786 & & 7.01\%\ & \textit{USDC}-\textit{WETH}, \textit{WETH}-\textit{WETH}, \textit{WETH}-\textit{USDC} & Uniswap $\times$ 2, SushiSwap & & 0.0183 & 0.07 & & 421,438 & 30,510 & 99,968\\
    PoS & 2.51 (7.86\%) & 52.2\% & 87.7\% & 56 (148) & 87,035 & & 7.89\% & \textit{WETH}-\textit{USDC}, \textit{WETH}-\textit{WETH}, \textit{USDC}-\textit{WETH} & Metamask, Uniswap $\times$ 2 & & 0.0105 & 0.01 & & 0 & 6,014 & 7,486 \\
    \bottomrule
    \end{tabular}}
    \caption{Findings comparison between PoW and PoS.
    % \textit{Count (m/m)} represents the count of private transactions in millions and its percentage per month.
    % \textit{Distribution} represents the percentage of blocks having no more than 10/20 private transactions.
    % \textit{Position} represents the average position of private transactions among the average number of transactions in a block. 
    % \textit{Fail (0 tip)} represents the count of failed private transactions with the least transaction cost, giving 0 tip to block creators.
    % \textit{MEV Bot} represents the percentage of private transactions calling MEV Bots for MEV extraction.
    % \textit{Mining Profits (\textit{ETH})} represents the average profits earned by block creators per type per month: Type I refers to the block rewards; Type II refers to the profits from private transactions; Type III refers to the profits from normal transactions.
    }
    \label{tab:findings} 
    \vspace{-20pt}
    \end{center}
\end{table*} 

It has been over three years since \pri \txs were first introduced by \textit{SparkPool} in August 2020~\cite{first-ptx}.
Several works have analyzed the Ethereum blockchain and DeFi platforms,
but they mainly focus on 
1) detecting bugs from smart contracts \cite{krupp2018teether,frank2020ethbmc,torres2021confuzzius},
2) measuring Ethereum networks and \txs \cite{lee-measure,chen2020understanding,Zhao2021TemporalAO}, and
3) analyzing \txs to uncover attacks \cite{Zhang2020TXSPECTORUA,wu2021defiranger,sereum-ndss19}.
To date, there are only a few papers~\cite{piet2022extracting,weintraub2022flash,qin2021quantifying,capponi2022evolution,yang2022sok,lyu2022empiricalstudyethereumprivate} that touched upon \pri~\txs as by-products when studying MEVs, mainly in PoW Ethereum. 
However, the scope of \pri~\txs goes beyond MEVs. A complete view of private \txs such as their impacts on the Ethereum ecosystem remains unclear. Moreover, since Ethereum has transitioned into PoS recently, the status of \pri \txs in PoS Ethereum is unknown. 

\subsection{Key Research Questions}
In this paper, we make the \textit{first} step towards understanding \pri~\txs and their impacts on the Ethereum ecosystem in both PoW and PoS.
We collect two datasets for our analysis, including a 15.5-month PoW dataset from June 1, 2021 to September 15, 2022 and a 15.5-month PoS dataset from September 15, 2022 to December 31, 2023, from reliable public data sources~\cite{Blocknative,etherscan}. 
This paper aims to answer three key research questions:
\begin{packeditemize}
    \item What are the transaction characteristics of private transactions, and how do they differ between PoW and PoS?
    % Transaction characteristics encompass various aspects related to the behaviors of private transactions. Our objective is to explore the trends and fundamental characteristics of private transactions on-chain, paying particular attention to the variances between PoW and PoS.
    
    \item What are the usages of private transactions, and how do they differ between PoW and PoS?
    % Private transactions serve as a means to avoid attacks and are often utilized for extracting MEV. However, the practical usages of private transactions, particularly the distinctions between PoW and PoS, remain unclear. Our objective is to uncover the mysteries surrounding the practical usages of private transactions and their variations between PoW and PoS.

    \item What are the monetary impacts of private transactions, and how do they differ between PoW and PoS?
    % Transaction users are required to pay transaction fees as rewards to block creators. Beyond these fees, private transactions may incur additional costs, potentially resulting in increased rewards for block creators. Our objective is to understand the monetary flow from users to block creators, with a specific focus on the disparities between PoW and PoS.
\end{packeditemize}

\subsection{Findings}
We summarize our findings in \tabref{tab:findings}, with a specific focus on the comparison between PoW and PoS Ethereum. 

\bheading{Transaction characteristics (\secref{sec:dyn}).}
% To show the overall characteristics of \pri \txs, we evaluate the count, distribution, position, and failure of \pri \txs.
We find that the monthly percentage of private transactions increases slightly from 2.45\% in PoW Ethereum to 7.86\% in PoS Ethereum, with a significant rise in the count from 0.95 million to 2.51 million per month. Additionally, the count of private transactions per block also increases in PoS Ethereum. This increase is likely due to reduced transaction costs and a growing interest in private transactions.
One noteworthy change in PoS Ethereum is the positioning of private transactions within mined blocks. Unlike in PoW Ethereum, where they are consistently placed at the block's beginning (position 16 in 224 transactions per block on average), in PoS Ethereum, they are now positioned at 56 out of 148 transactions per block on average. However, their functionalities remain unchanged.
Furthermore, we observed a higher adoption of the 0 tip strategy in case of transaction failure in PoS Ethereum (15,709 cases) compared to PoW Ethereum (635 cases). Using this strategy, private transactions make direct payments to block creators rather than paying high transaction fees as tips for prioritization. In the event of a transaction failure, there is no loss (0 tip), as the direct payment is reverted. In contrast, transaction fees are charged regardless of success or failure. Therefore, we suggest private transaction users employ this strategy.

\bheading{Transaction usages (\secref{sec:app}).}
% To study usages of private transactions, we quantify the entities (senders, receivers, and function calls), as well as the frequently used tokens and DeFi applications.
We find that while the involved active addresses differ, \pri \txs are mainly used for 1) MEV extraction, 2) money transfer, and 3) calling well-known DeFi services, in both PoW and PoS. 
Taking money transfer as an example, miners (e.g., \textit{Ethermine}) usually send \pri \txs to redistribute mining profits in PoW. 
However, using private transactions for MEV extraction has significantly decreased in PoS Ethereum. This shift is evident as the percentage of private transactions involving MEV Bots has increased from 14.02\% in PoW Ethereum to 17.59\% in PoS Ethereum.
We also find that the most commonly used token pairs and DeFi applications are largely consistent between PoW and PoS Ethereum. For example, the top three exchange token pairs involve tokens \textit{WETH} and \textit{USDC}, and the most frequently used DeFi applications are primarily Uniswap. This suggests that private transaction users predominantly engage with well-established DeFi applications for MEV extraction, often keeping stable coins (e.g., \textit{USDC}) and \textit{WETH} as profits.
We also find that private transactions are utilized in DeFi attacks, with an increased prevalence observed in PoS Ethereum compared to PoW Ethereum.
Despite the intended purpose of protecting users from attacks, private transactions can still leave users vulnerable to MEV-related risks and DeFi attacks. Therefore, we recommend that users consider exchanging a small number of tokens more frequently to reduce their exposure to MEV seekers.

\bheading{Transaction monetary impacts (\secref{sec:eco}).}
% To reveal the monetary flow behind \pri \txs, we evaluate the transaction cost from the user's perspective and the mining profits from the block creator's perspective.
We find that private transactions in PoS Ethereum offer cost advantages over those in PoW Ethereum, encompassing both transaction fees (0.0081 \textit{ETH}) and direct payments (0.01 \textit{ETH}), which are notably lower compared to PoW Ethereum's transaction fees (0.07 \textit{ETH}) and direct payments (0.0183 \textit{ETH}).
This cost reduction in PoS Ethereum can be attributed to the increased diversity of \pri \tx service providers. In PoW Ethereum, Flashbots monopolizes the market, mining 82.9\% of blocks containing \pri \txs. However, in PoS Ethereum, we observe the emergence of more service providers, such as \textit{BloXroute}, leading to Flashbots holding only a 41.6\% market share (for detailed information, refer to \S\ref{app:eco:par}).
Moreover, the profits of block creators in the PoS era experience significant reductions due to the absence of block rewards (resulting in 0 \textit{ETH} profits per month), decreased tips from private transactions (amounting to 6,014 \textit{ETH} profits per month), and diminished tips from normal transactions (totaling 7,486 \textit{ETH} profits). This is in stark contrast to PoW Ethereum, where the corresponding figures stand at 421,438 \textit{ETH}, 30,510 \textit{ETH}, and 99,968 \textit{ETH}, respectively.
The consensus mechanism transition and the subsequent rise of diverse \pri \tx service providers offer multiple advantages to users, including faster transaction processing times and reduced transaction costs. Furthermore, this evolution contributes to a less centralized Ethereum blockchain to some extent.
\section{Background}
\label{sec:bg}

% Ethereum~\cite{Buterin2013} stands as a public, decentralized, and permissionless blockchain platform. In Ethereum, nodes, often referred to as Ethereum clients, are interconnected through a P2P network for effective data communication, such as transaction broadcasting.

% \subsection{Transaction Basics}
% \subsubsection{Transaction Entities.} 
% Ethereum has two types of accounts: Externally Owned Accounts (EOAs) and Smart Contracts. EOAs are devoid of code and function as standard cryptocurrency wallets, while smart contracts represent Turing-complete, self-executing programs that run on the Ethereum blockchain.

% \bheading{Senders and receivers.}
% Transactions always originate from an EOA (sender) and are directed towards a specific account (receiver). When the receiving account is also an EOA, the transaction constitutes a straightforward transfer between two accounts. However, if the receiving account is a smart contract, the Ethereum Virtual Machine (EVM) executes the related function call within the contract.

% \bheading{Function calls.}
% The function to be executed is determined by the first four bytes of the call data in a function call. For example, the function signature \textit{0xa9059cbb} corresponds to the function \textit{transfer(address to, uint256 value)}. If the execution completes successfully without errors, the transaction is considered successful. Conversely, if any issues arise during the transaction, all operations are reversed, and the transaction fails.

Ethereum stands as a public, decentralized, and permissionless blockchain platform. It has two types of accounts: Externally Owned Accounts (EOAs) and Smart Contracts. Transactions always originate from an EOA.

\subsubsection{Transaction Cost}
The cost associated with a transaction consists of the transaction fee and the direct payment.

\bheading{Transaction fee.}
Every transaction is obligated to pay a fee to cover the computational resources required for execution. This fee is determined by both the gas used and the gas price, and it is calculated as follows: \textit{TxFee = UsedGas × GasPrice}. Here, \textit{UsedGas} signifies the amount of gas utilized during \tx execution, and \textit{GasPrice} represents the user's chosen payment per unit of gas.

Prior to the implementation of EIP-1559~\cite{eip1559}, \textit{GasPrice} could be set to 0. However, following its activation on August 5, 2021 (during the PoW era), \textit{GasPrice} is composed of two components: the base fee, algorithmically determined for each Ethereum block, and the priority fee, which is an optional incentive for block creators to include \txs in newly mined blocks. 

\bheading{Direct payment.} 
Aside from transaction fees, private transactions typically involve a direct payment to block creators in the form of tips. Generally, the more substantial the tips offered by private transactions, the higher the likelihood of their swift inclusion in desired positions (e.g., being mined before the victim transaction in a front-running attack).

\subsubsection{Transaction Usages}
Transactions on Ethereum serve multiple purposes, including interaction with DeFi applications for token exchanges and Maximum Extractable Value (MEV) extractions to gain profits.

\bheading{Tokens and DeFi applications.} Every Ethereum user has the ability to possess and generate tokens. The Ethereum Request for Comments 20 (ERC20)~\cite{erc20} stands as the most prominent token standard, specifically designed for fungible tokens. Among these tokens, stablecoins adhere to the ERC20 standards to maintain price stability. For instance, Tether represents a stablecoin with a value pegged to 1 USD.

DeFi applications are financial applications constructed on the foundation of smart contracts. These applications are primarily geared towards facilitating financial activities such as borrowing, lending, token swapping, asset management, and a wide array of other financial functions. Notable examples of DeFi applications include Uniswap and SushiSwap.

\bheading{MEV and MEV Bots.}
The concept of MEV was first introduced by \textit{Daian et al.}~\cite{flashboy}. 
MEV, while sometimes referred to as miner extractable value, is predominantly pursued by users rather than block creators. Users actively seek MEV opportunities and share the profits earned with block creators. Block creators, in this context, simply need to incorporate MEV-related transactions into mined blocks and receive a portion of the shared profits from users.

To compete for MEV extraction, MEV Bots have been developed to automatically monitor the blockchain network for submitting transactions with potential MEV opportunities. These bots swiftly initiate related transactions via private transaction service providers, such as Flashbots~\cite{flashbot}. 
Since its inception in 2019, MEV has been the subject of intensive research~\cite{piet2022extracting,weintraub2022flash}. The extensive exploration of MEV may introduce instabilities and reorganizations on the blockchain network because block creators are incentivized to re-mine blocks. Moreover, users may face financial losses if they are targeted or front-run by MEV-related transactions.

% \begin{figure*}[t]
% \centering
% \includegraphics[width=0.8\linewidth]{figs/sigmetrics-bg2.pdf}
% \caption{
% The lifecycle of a normal transaction (\texttt{Tx}) vs. a \pri \tx (\texttt{PTx}) in Ethereum. 
% User \texttt{A} sends a \texttt{Tx} to a node ($N_1$), which is then broadcast to others, such as $M/V$ nodes (M represents the miner in PoW and V represents the validator in PoS).  
% However, the workflow is different for a \texttt{PTx} sent from User \texttt{B}, in which \texttt{PTx} is sent directly to $M/V$ via the \pri \tx service provider.
% Providers in PoW Ethereum only consist of relays that forward bundles of \pri \txs to miners; instead, in PoS Ethereum, relays in providers choose the best blocks from block builders and forward them to MEV-Boost, which picks the most profitable blocks to validators.
% }
% \label{figs:bg}
% \end{figure*}

\subsection{Private Transactions} 
\label{sec:bg:pri}
Private transactions are a special category of transactions designed to operate discreetly without the typical broadcasting on the public P2P network. In Ethereum, both private and normal transactions are processed and incorporated into blocks by block creators. However, changes in Ethereum's consensus mechanism from PoW to PoS have impacted the lifecycle of private transaction mining. 

\bheading{Similarities.}
In general, there are three common stages for both \pri  and normal \txs as follows.

\begin{packeditemize}
\item Stage 1: initialization. Users construct \txs and submit them to Ethereum. 
\item Stage 2: mining. If transactions are successfully added to blocks, they are recorded on the blockchain, in which their positions are typically based on the profitability of the transactions.
\item Stage 3: execution. Once recorded on the blockchain, transactions are executed and their results are determined.
\end{packeditemize}

\bheading{Differences.} 
Besides the similarities,
there are two distinct differences between private and normal transactions. 

\textit{First}, normal \txs are sent to Ethereum nodes, while \pri \txs are sent directly to the miners in PoW Ethereum or validators in PoS Ethereum, via \pri \tx service providers. 
In PoW Ethereum, users first submit \pri \txs. 
Then bundle propagation services (relays)  pack \pri \txs into bundles and forward the bundles to miners. 
Finally, miners mine the \pri \txs into the front places of the new mined block 
as long as they are satisfied with the rewards.
However, in PoS,
after the users send \pri \txs, 
block builders take \pri \txs and fetch normal \txs from public network, to construct new blocks;
every block builder generates the most optimal block and forwards it to it connected relays. 
After that, every relay chooses the best blocks and forwards it to the validators.
% MEV-Boost~\cite{mev-boost}, which then selects the block with most profits for validators. In particular, MEV-Boost is an implementation of proposer-builder separation (PBS)~\cite{pbs} built by Flashbots. 

\textit{Second}, there is a price of being a \pri \tx, which is usually to pay the additional fee by directly sending money during the execution of \tx.
In PoW Ethereum, \pri \tx senders directly make payments to miners; while in PoS Ethereum, \pri \tx senders make payments to block builders first and then block builders collect all the payments to validators.

\vspace{-8pt}
\section{Dataset}
\label{sec:dataset}
To empirically study \pri \txs, we extract the necessary data from reliable data sources, such as Blocknative~\cite{Blocknative} that has been commonly used by other research~\cite{yang2022sok,adams2023costs, heimbach2023ethereum, chan2017ethereum, yuan2020detecting, chen2020phishing}.

\subsection{Transaction Dataset}
\label{sec:dataset:tx}
We collect the necessary \tx and block information from Blocknative~\cite{Blocknative} and Etherscan~\cite{etherscan}.
1) For each \tx, we collect the basic information, including transaction {hash}, block number, success status, sender, receiver, \textit{ETH} value, input, used gas, gas price, and \tx fee. We also fetch money flows inside \txs, consisting of token sender, token receiver, token amount, and token address. 
2) For every block, we obtain block number, block creator, block reward, and base fee. 

\bheading{15.5-month PoW transaction dataset.}
We collect a 15.5-month Proof of Work (PoW) dataset spanning from June 1, 2021 (block 12,545,219) to September 15, 2022 (block 15,537,394), encompassing a total of 556,270,468 transactions. Within this dataset, we identified 14,810,392 transactions (2.66\% of the total) as private transactions based on the dataset provided by Blocknative~\cite{Blocknative}, while the remaining 541,460,076 transactions were classified as normal transactions. 
% We source private transactions from Blocknative~\cite{Blocknative}, a platform also utilized in prior studies~\cite{adams2023costs, heimbach2023ethereum}. 

Specifically,  Blocknative employs a network of Ethereum nodes worldwide to detect private transactions; if a transaction is not observed by any node but is still confirmed in a block, it is classified as a private transaction. This approach has also been employed by \textit{Sen} et al.~\cite{yang2022sok}. Due to resource constraints for extensive, multi-node monitoring over extended periods, we chose to utilize the dataset provided by Blocknative. To validate the accuracy of the Blocknative private transaction datasets, we manually verified a subset of 100 randomly selected private transactions, all of which were found to be 100\% accurate. However, occasional mis-classifications of private transactions due to network latency may occur, which we acknowledge as a limitation. 
% (see ~\secref{sec:dis} for discussing private transaction datasets).
% Specifically, Blocknative employs a network of global Ethereum nodes to monitor transactions, categorizing those not initially observed by any node but later confirmed in blocks as private transactions. A similar methodology is utilized by \textit{Sen} et al.~\cite{yang2022sok}. Due to limitations in resources for extensive, multi-node monitoring over prolonged periods, we opt to utilize the dataset provided by Blocknative. To ensure the accuracy, Blocknative operates multiple global nodes for transaction monitoring.

\bheading{15.5-month PoS transaction dataset.}
To compare \pri \txs in PoW with those in PoS Ethereum, we collect another 15.5-month PoS dataset spanning from September 15, 2022 (block 15,537,394) to December 31, 2023 (block 18,908,894), comprising a total of 499,327,807 transactions. According to Blocknative~\cite{Blocknative}, this dataset includes 30,062,232 private transactions (6.02\% of the total), with the remaining 469,265,575 transactions categorized as normal transactions.

\subsection{Smart Contract Label Dataset}
\label{sec:dataset:sc}
To measure the behaviors of different entities in \pri \txs, we check whether they have the following labels; if so, we collect the corresponding information from Etherscan Label Cloud:
 
 \iheading{1) MEVBot}: addresses whose label are MEVBots. For every unique sender and receiver of \txs in our dataset, we check its label from Etherscan. If the label is MEVBot, we collect it.
 
 \iheading{2) DeFi}: DeFi address and related label. Similarly, we collect the addresses which are marked as DeFi by Etherscan, and their labels. 
 
 \iheading{3) Token}: For each smart contract labeled as ERC20 token, we obtain the token address and name from Etherscan, and historical price data from TradingView~\cite{tradingview}.

% \vspace{-8pt}
\section{Transaction Characteristics}
\label{sec:dyn}
To give an overview of transaction dynamics, we measure the basic characteristics of private transactions and the failed private transactions that are mined but not successfully executed.

\subsection{Basic Characteristics}

\subsubsection{Count and Percentage} 
\begin{figure}[t]
    \centering
    \begin{subfigure}[b]{.48\linewidth}
        \centering
        \includegraphics[width=1.0\linewidth]{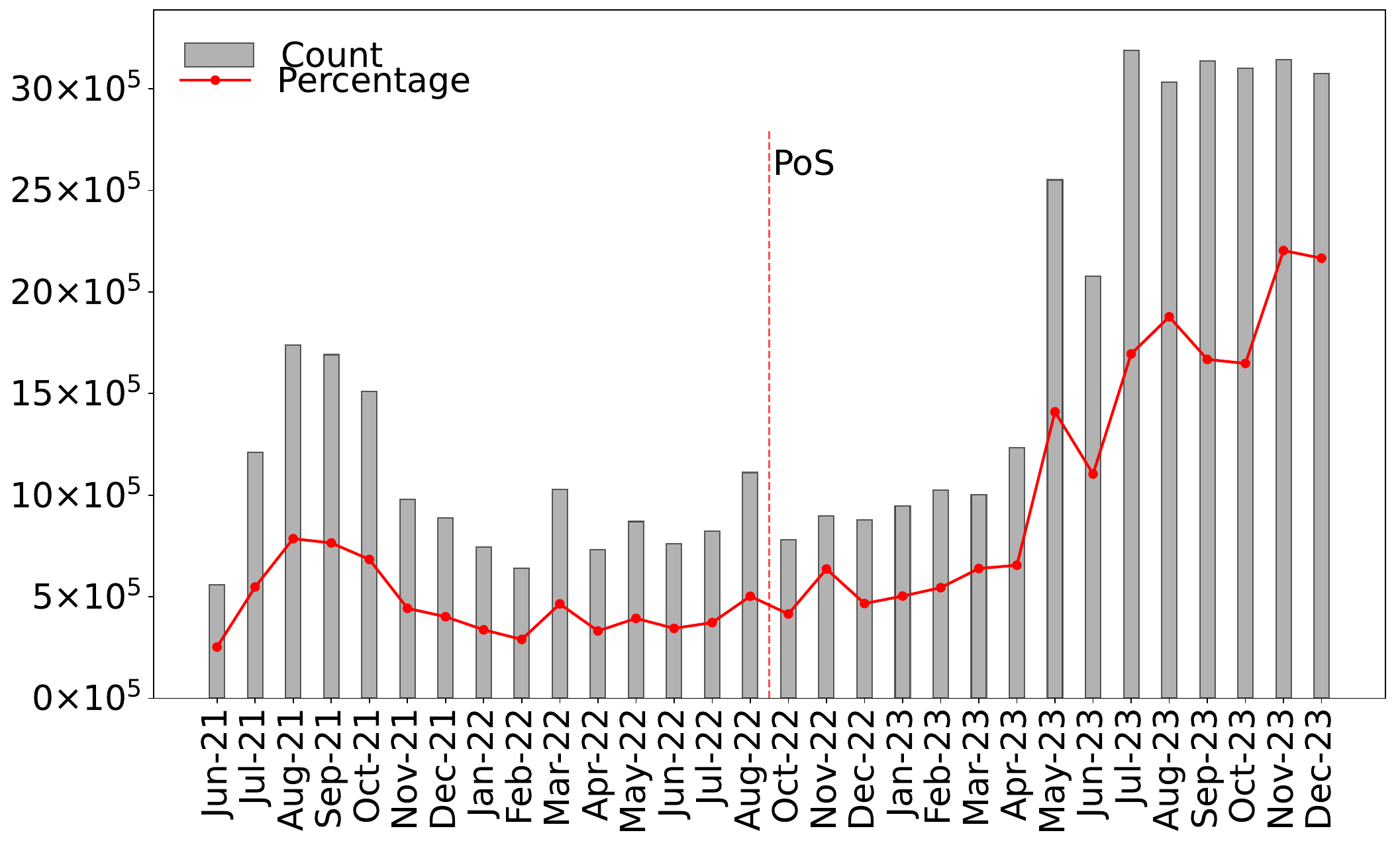}
        \caption{Count per month}
        \label{figs:txnum}
    \end{subfigure}
    \hfill
    \begin{subfigure}[b]{.48\linewidth}
        \centering
        \includegraphics[width=1.0\linewidth]{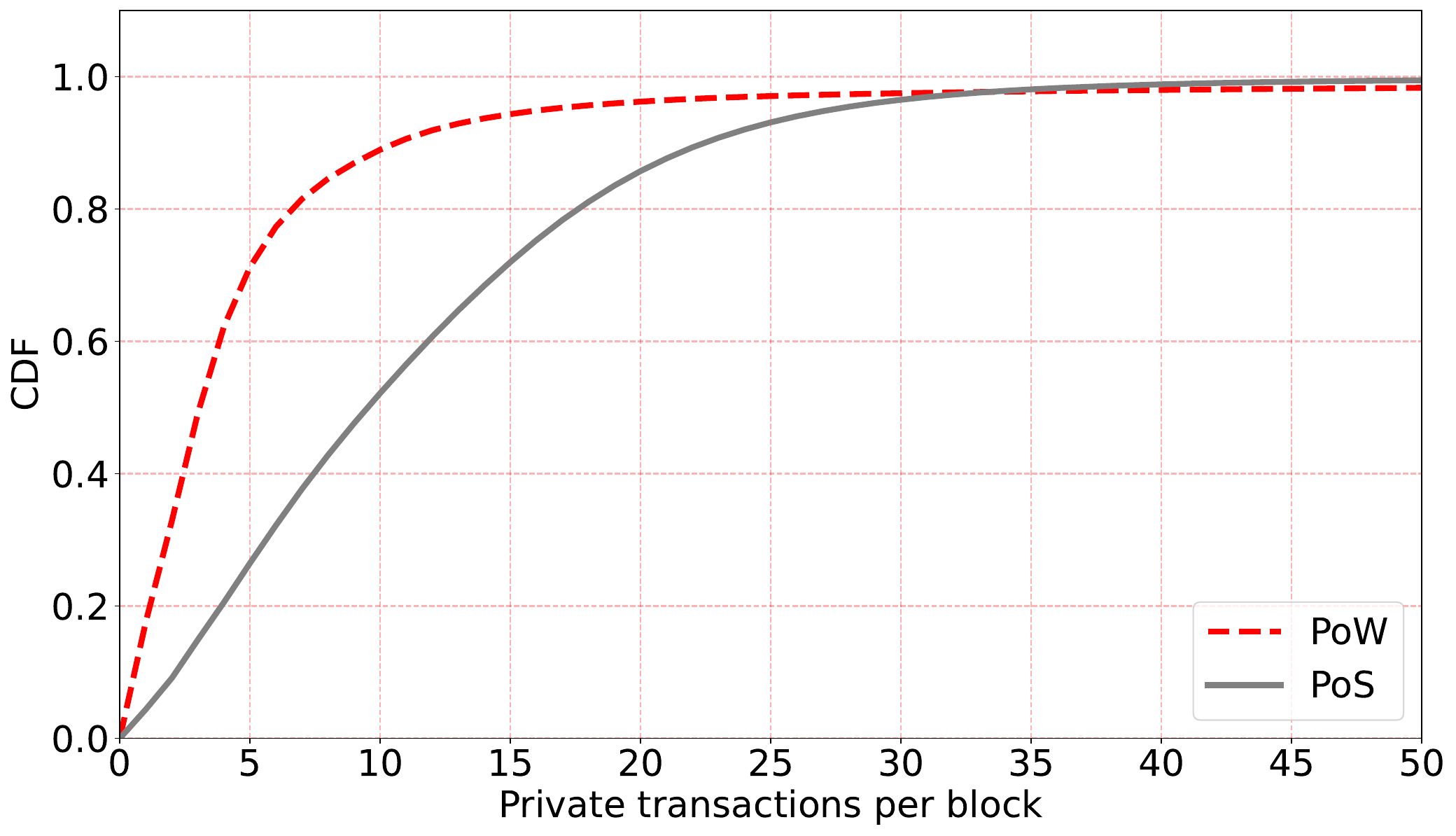}
        \caption{Count per block}
        \label{figs:rate_mix_perblock}
    \end{subfigure}  
    \caption{Count of private transactions per month/block.}
    \vspace{-10pt}
\end{figure}

\figref{figs:txnum} shows the count and percentage of \pri \txs per month in both PoW and PoS Ethereum.

We observe that the quantity of private transactions during the PoS era surpasses that of the PoW era, with a gradual increase in the monthly count of private transactions during the PoS era, except for May 2023.
In PoW Ethereum, there are a minimum of 0.56 million private transactions each month, averaging around 0.95 million private transactions monthly. These private transactions make up an average of 2.45\% of all transactions each month.
Conversely, in PoS Ethereum, the number of private transactions consistently remains at a minimum of 0.78 million per month, averaging approximately 2.51 million private transactions monthly, constituting 7.86\% of the total transactions. 
In comparison, PoS Ethereum showcases a notable increase in both the absolute number and proportion of private transactions, indicating a shifting trend towards privacy within Ethereum blockchain network.

\subsubsection{Distribution}
\label{sec:char:dis}
\figref{figs:rate_mix_perblock} illustrates the Cumulative Distribution Function (CDF) of private transaction count per block. This data is derived from blocks containing at least one private transaction, totaling 1,814,592 blocks (60.64\% of the total) in the PoW era and 2,500,979 blocks (74.18\%) in the PoS era.

We observe that in PoW Ethereum, 88.92\% of these blocks contain fewer than 10 private transactions, and 96.35\% of them feature no more than 20 private transactions. In PoS Ethereum, 52.26\% of blocks have less than 10 private transactions, and 87.71\% of blocks contain no more than 20 private transactions.
Moreover, there are more blocks in the PoS era than in the PoW era. This phenomenon can be attributed to the PoS era generating a greater number of blocks within a single month, as block mining time is reduced to 12 seconds in the PoS era, compared to 13-14 seconds in the PoW era~\cite{pow-time}.
In comparison, we observe a notable shift towards a higher proportion of blocks with more private transactions in the PoS era, indicating potentially enhanced privacy features.

\subsubsection{Position}
The position denotes the sequential order of a transaction mined within a block. For example, if a private transaction has a position of 2 (indexing from 0) in block 100, it means that this private transaction was mined in block 100, and there are two transactions preceding it within the block.

We observe significant differences in the positioning of private transactions between the PoW and PoS eras. In PoW Ethereum, our analysis covers 1,814,592 blocks containing at least one private transaction, with an average of 186 transactions per block. Interestingly, we find a consistent pattern where all private transactions are uniformly situated at the top of the blocks, typically occupying position 16. This pattern suggests that miners in the PoW era tended to prioritize and group private transactions together, likely due to the competitive environment for block space alongside other transactions.

On the other hand, during the PoS era of Ethereum, we examine 2,500,979 blocks with at least one private transaction, revealing an average of 148 transactions per block, with the average position of private transactions being 56. Notably, the average position of private transactions shifts to position 56. Furthermore, we find that only 44.77\% of all private transactions are concentrated within the top 10 positions, while 57.93\% are distributed across the top 20 positions. This indicates a more equitable distribution of private transactions within PoS-era blocks. As an illustration of this dispersed arrangement, consider block 15,843,819, where private transactions were situated in positions 1-4, 6-10, 11-18, and 219, totaling 219 transactions.

% \lesson{While the positioning of private transactions in PoS Ethereum is notably further back, their functionalities remain unchanged. In PoW Ethereum, private transactions typically occupy the beginning of blocks, with an average position of 16 (186 transactions per block on average), primarily clustered toward the block's start. Conversely, in PoS Ethereum, the average position shifts to 56 (148 transactions per block on average), with private transactions being more scattered within the blocks.}

\subsection{Failed \Pri \Txs} 
When a private transaction fail, its execution is rolled back, resulting in the cancellation of the direct payment associated with the transaction. However, users are still obligated to cover the transaction fee. Consequently, it is advisable for users to consider employing a 0 tip strategy in case of failure. This strategy involves making direct payments instead of opting for a higher transaction fee.

We have observed a significant increase in the number of failed private transactions in the PoS era. Some users have taken a cautious approach by setting the gas price to 0 before EIP-1559 or the priority fee to 0 after EIP-1559 to save on costs in case transactions fail. In PoW Ethereum, we identified a total of 42,950 (0.29\%) failed private transactions. Before EIP-1559, there were 1,257 failed private transactions, with 68 of them having a gas price (and tip) of 0. After EIP-1559, the number of failed private transactions rose to 41,693, with 2,718 (6.52\%) having a priority fee of 0. In PoS Ethereum, every private transaction requires payment of the base fee as mandated by EIP-1559. We recorded 658,362 (2.20\%) failed private transactions, with 13.22\% of them featuring a priority fee (tip) of 0.

% \lesson{
% In PoS Ethereum, there has been a notable surge in the occurrence of failed private transactions with a 0 tip, contrasting with PoW Ethereum. This increase can be primarily attributed to PoS-era private transaction service providers, such as Flashbots, actively promoting the use of 0 priority fees as a strategy to conserve transaction fees in the event of transaction failures~\cite{why-p0}.
% }

\section{Transaction Usages}
\label{sec:app}
To offer a comprehensive perspective on the usages of private transactions, this section initiates by assessing the entities involved, namely senders and receivers, along with their corresponding function calls. Moreover, given that private transactions are primarily employed for interactions with tokens and DeFi applications, we also measure the popularity of tokens and the usage of prominent DeFi applications. Additionally, we observe that private transactions are utilized for launching DeFi attacks, aiming to evade monitoring by whitehats and rescue attempts.

\subsection{Entities}

\subsubsection{Senders and Receivers}
To give an overview, we study the top 10 senders of private transactions and top 10 receivers.
% (see \appref{app:en:sender}) and top 10 receivers (see \appref{app:en:receiver}).

We can classify the top 10 senders into different categories to study the \pri \tx usages.
In PoW dataset, top 10 senders account for 10.6\% private transactions. We have 2 miners (\textit{Ethermine}, \textit{F2Pool}), 7 EOAs that mainly interact with MEV Bots to search for MEV opportunities, and 1 DeFi application related address (\textit{Alameda}). 
In PoS dataset, top 10 senders account for 5.11\% private transactions. We have 2 block builders (\textit{Flashbots: Builder}, \textit{Builder0x69}) who build blocks for validators to propose and then forward mining profits to validators, 7 EOAs, and the miner (\textit{Ethermine}) which is involved in money transfers. 

We can also classify the top 10 receiver into different categories.
 % (see~\tabref{tab:top_receivers})
In PoW dataset, the \pri \txs received by top 10 receivers relate to 15.37\% private transactions. 5 MEV Bots are involved in 7.01\% of private transactions, 4 DeFi application related addresses (\textit{Uniwswap V2/V3, SushiSwap}) in 7.21\%, and 1 token address (\textit{Tether USDT}) in 1.15\%. In PoS dataset, the \pri \txs calling top 10 receivers relate to 11.85\% \pri \txs. 5 MEV Bots are involved in 7.89\% of private transactions, 4 DeFi application addresses (\textit{Uniwswap V2/V3, Metamask:Swap, Jump Trading 2}) in 3.22\%, and 1 token address (\textit{WETH}) in 0.74\%.

\subsubsection{Function calls}
We study the top 10 function calls sorted by the count of related private transactions.
% (see \appref{app:en:func}). 

We observe that there are some overlaps between the top 10 function calls in PoW and PoS eras. First, the top 1 function signature is 0x in both eras, which represents the empty call to EOA addresses or the fallback() function to smart contracts. Second, there are some token standard functions (\textit{transfer}) and DeFi services (\textit{execute, multicall}) widely used in both eras. Third, we find that although the function names are different, the top 10 function calls are mainly involved in token exchanges, such as the function \textit{swapExactTokensForTokens}. 

% \lesson{In PoW Ethereum, private transactions are primarily used for: 1) extracting MEV, 2) facilitating monetary transfers, such as miners distributing profits to mining nodes, and 3) interacting with well-established DeFi applications. In the transition to PoS Ethereum, the usage of private transactions remains somewhat analogous, but there is a noteworthy transformation in the most active addresses. 
% Furthermore, the role of private transactions in extracting MEV increases from 7.01\% in PoW Ethereum to 7.89\% in PoS Ethereum, as demonstrated by the top 5 MEV Bots.} 

\subsection{Tokens and DeFi Applications}

\subsubsection{Token Usage}
To provide an overview of token usage within private transactions, we examine the types of tokens involved and identify the most frequently used token exchange pairs. Additionally, we note a slight increase in the number of token flows.

\bheading{Token types.}
We observe that private transactions in PoS Ethereum involve a slightly higher variety of token types, implying increased complexity in PoS-era private transactions.

In PoW Ethereum, private transactions that exclusively involve the exchange of one, two, three, or more than three types of tokens account for 12.80\%, 69.40\%, 15.60\%, and 2.20\%, respectively. In PoS Ethereum, the corresponding percentages are 7.17\%, 72.28\%, 18.06\%, and 2.49\%. Comparing the PoS era to the PoW era, there is a marginal increase in the percentage of private transactions involving the exchange of two or more types of tokens. We suspect that these changes are influenced by the proliferation of DeFi applications and heightened token interoperability~\cite{defipos}.

\begin{figure}[t]
    \centering
    \begin{subfigure}[b]{.48\linewidth}
        \centering
        \includegraphics[width=1.0\linewidth]{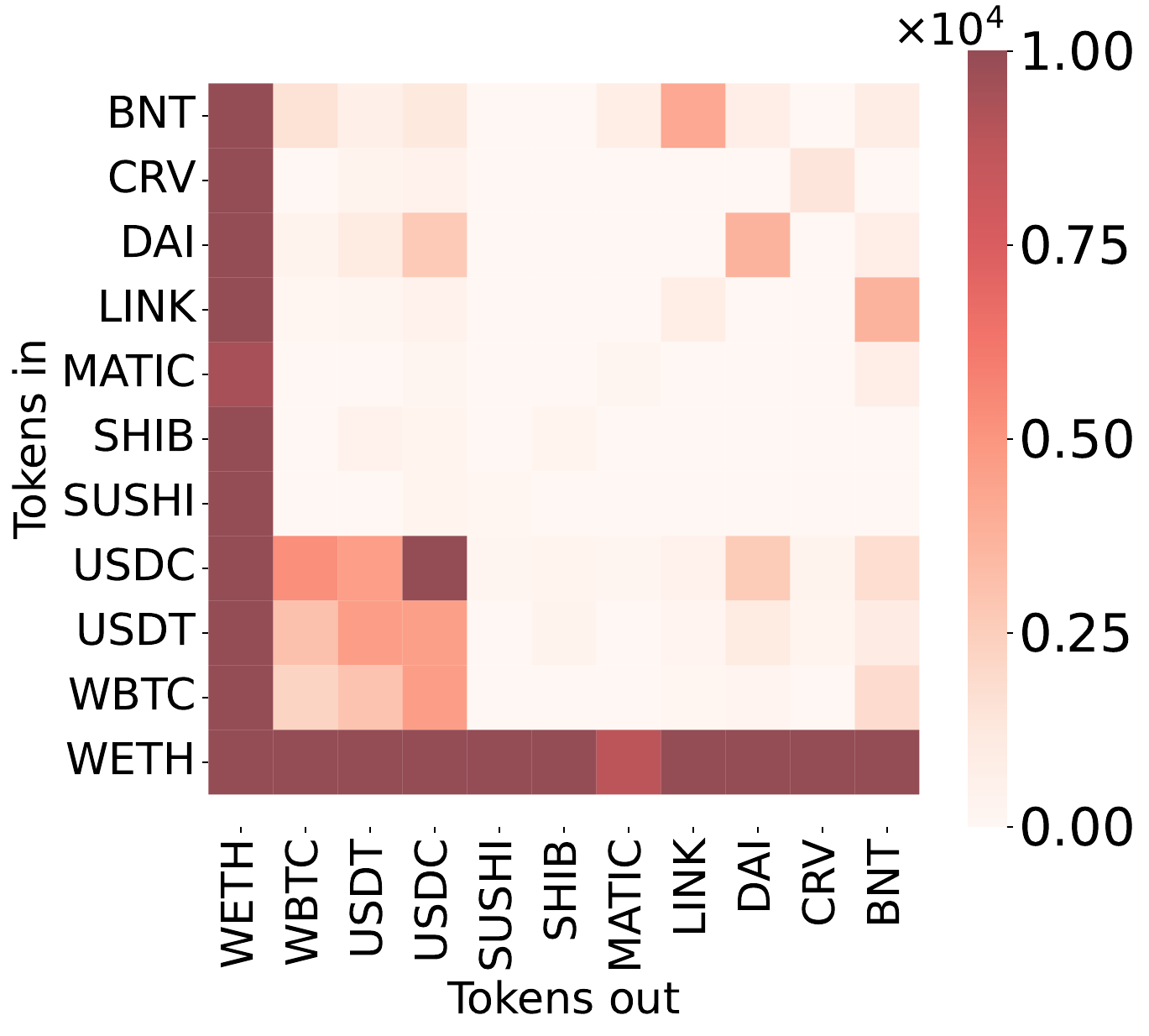}
        \caption{PoW}
        \label{figs:pairs}
    \end{subfigure}
    \hfill
    \begin{subfigure}[b]{.48\linewidth}
        \centering
        \includegraphics[width=1.0\linewidth]{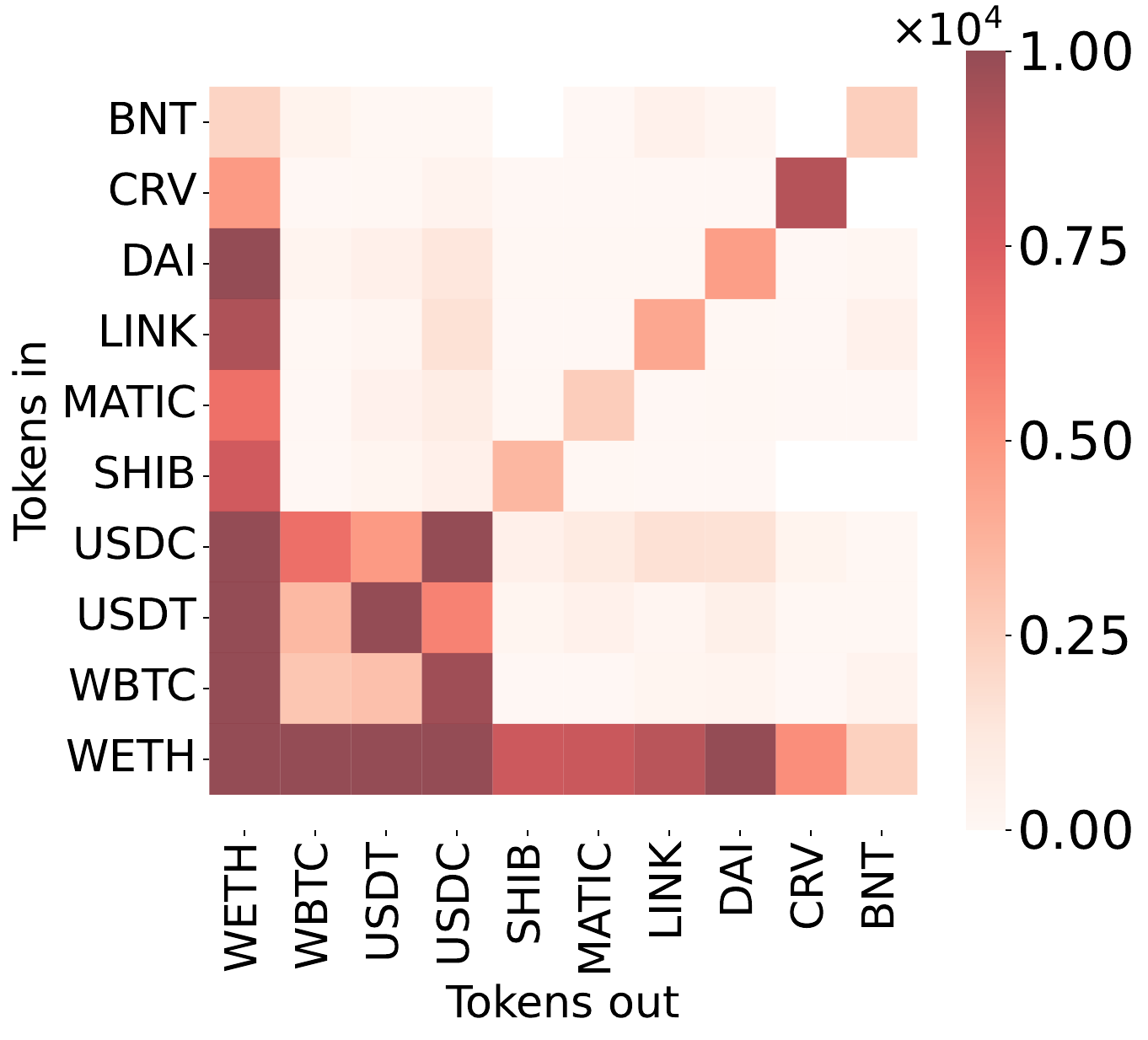}
        \caption{PoS}
        \label{figs:pos_pairs}
    \end{subfigure}
    \caption{Top token exchange pairs of \pri \txs.}
\end{figure}

\bheading{The most used token exchange pairs.}
\figref{figs:pairs} and \figref{figs:pos_pairs} display the leading token exchange pairs associated with private transactions for PoW and PoS data, respectively. 

We observe that the most frequently used tokens in both PoW and PoS eras are primarily stable coins (e.g., \textit{USDC}) and wrapped Ethereum (\textit{WETH}), which is the wrapped version of the native token \textit{ETH}. In the PoW era, the top three exchange pairs are \textit{USDC}-\textit{WETH} (2.22\%), \textit{WETH}-\textit{WETH} (2.08\%), and \textit{WETH}-\textit{USDC} (1.89\%). In the PoS era, the top three are \textit{WETH}-\textit{WETH} (3.59\%), \textit{USDC}-\textit{WETH} (2.67\%), and \textit{WETH}-\textit{USDC} (2.58\%).
Furthermore, we observe a significant increase in exchanges involving tokens with themselves in the PoS era compared to the PoW era. We suspect that this trend is driven by the growing popularity of yield farming~\cite{postoken2}.

\begin{figure}[t]
\centering
\includegraphics[width=.9\linewidth]{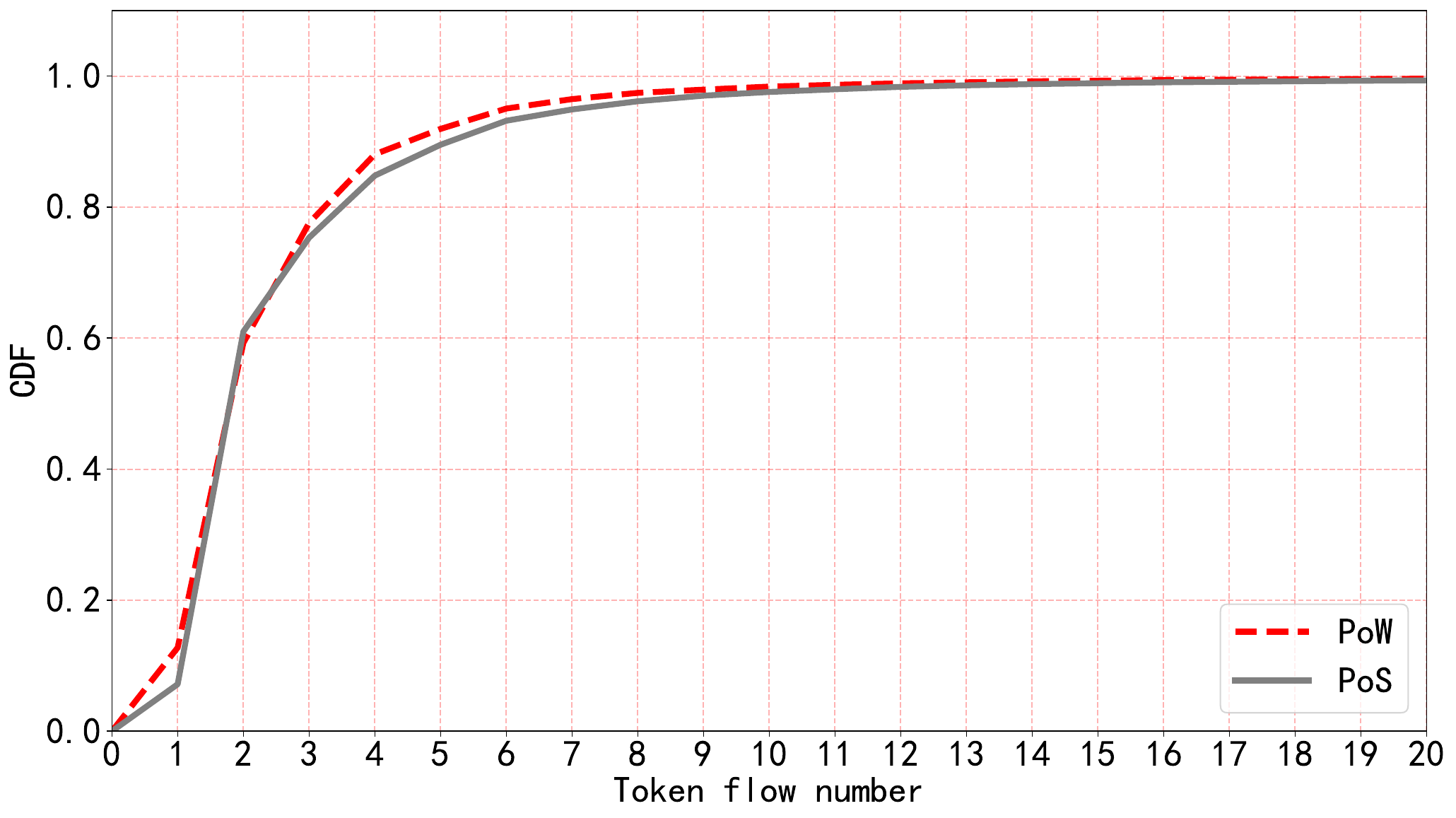}
\caption{Token flow number per \pri \tx.} 
\label{figs:token_flow}
\end{figure}

\bheading{Token flows.}
\figref{figs:token_flow} presents the distribution of token flow number of \pri \txs that have at least one token flow in both PoW and PoS Ethereum. 

We observe that the token flow number per \pri \tx is slightly larger in PoS era than that of PoW era, indicating the \pri \txs are becoming a little more complex, while the percentage of \pri \txs having token flow decreases slightly. In PoW Ethereum, we have 77.6\% \pri \txs with at least one token flow. Around 88.1\% have four or fewer token flows and 99.5\% have 20 or fewer token flows. The private transaction~\texttt{0x470ceedf} has the highest number of token flows, which is 600. In PoS Ethereum, we have 67.24\% \pri \txs containing token flow. Approximately 84.9\% of the transactions have four or fewer token flows and 99.2\% have twenty or fewer token flows. The private transaction~\texttt{0x09168f21} with the largest number of token flows has 1,412 token flows.

\subsubsection{DeFi Application Usage}

We study the top 10 DeFi addresses frequently utilized in private transactions, arranged by the number of private transactions associated with them.
We observe that the most popular DeFi applications remain consistent across both PoW and PoS eras, albeit with updated addresses for the same applications. Additionally, some addresses belong to the same DeFi applications. In PoW Ethereum, both the 1st and 2nd addresses are linked to Uniswap, while both the 5th and 8th addresses are associated with 1inch. In PoS Ethereum, the top 2nd and 3rd addresses still pertain to Uniswap, and the 4th and 9th addresses belong to 1inch. Notably, 1inch has transitioned to V5 in the PoS era. Across both PoW and PoS eras, four DeFi applications stand out as widely utilized in private transactions: Uniswap, SushiSwap, 1inch, and Metamask.

% \lesson{Although private transactions exhibit a slight increase in complexity within PoS Ethereum, particularly in terms of the number of tokens transferred and token types per transaction, it's noteworthy that the top tokens involved, such as \textit{WETH} and the stablecoin \textit{USDC}, as well as popular DeFi applications, such as \textit{Uniswap}, remain consistent across both PoW and PoS Ethereum.}

\subsection{DeFi Attacks}
In our PoW dataset, we collected data on 189 DeFi attacks, while in our PoS dataset, we recorded 85 DeFi attacks from reliable sources, including DEFIYIELD Rekt Database~\cite{defiyield} and BlockSec Phalcon~\cite{blocksec}. In PoW Ethereum, 20 attacks (10.58\%) involved private transactions, whereas in PoS Ethereum, 35 attacks (41.18\%) utilized private transactions. This indicates a notable increase in the use of private transactions for DeFi attacks in the PoS era.

\begin{figure}[t]
    \centering
    \begin{subfigure}[b]{.48\linewidth}
        \centering
        \includegraphics[width=1.0\linewidth]{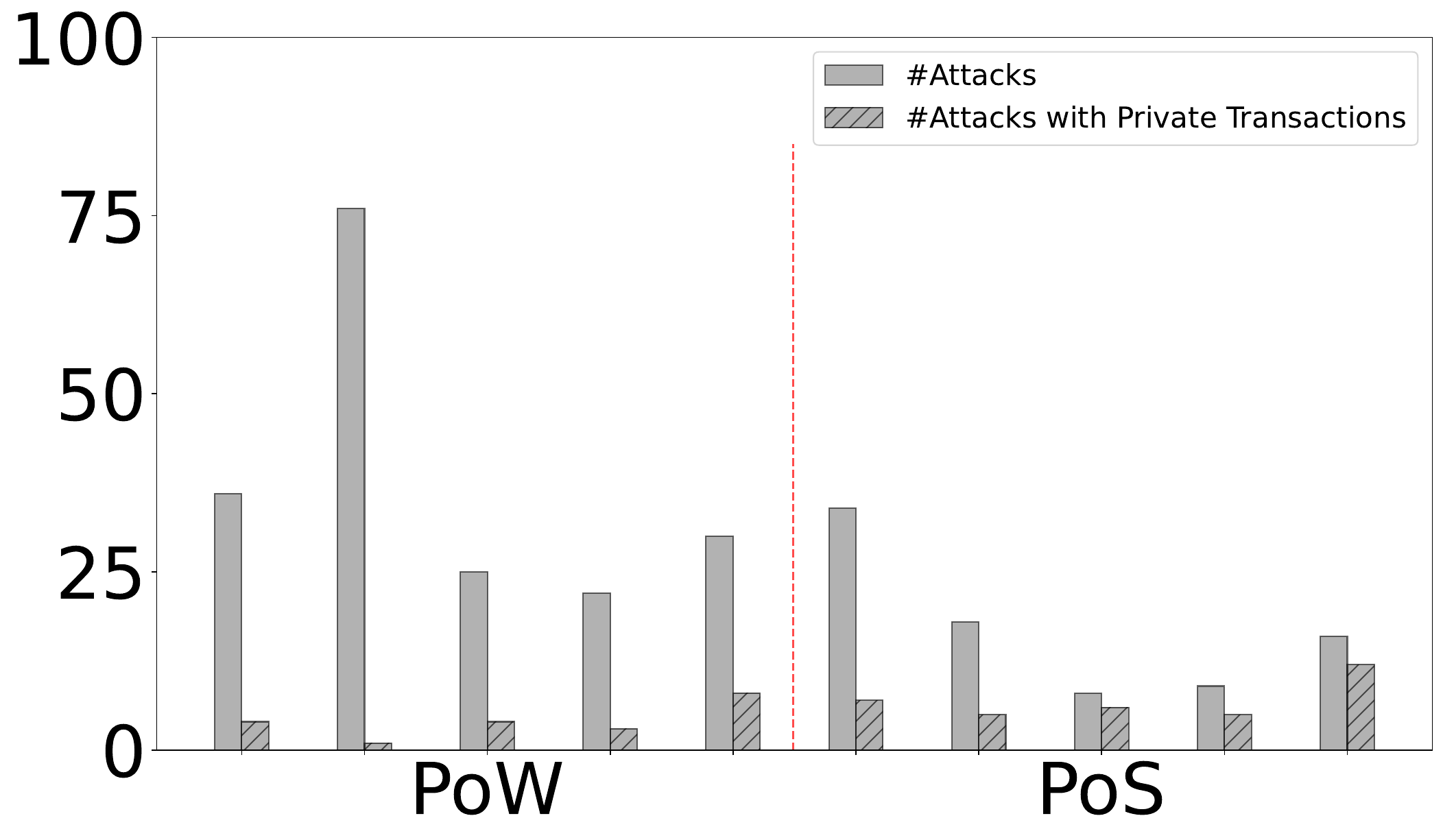}
        \caption{Count of DeFi attacks}
        \label{figs:defi_attacks}
    \end{subfigure}
    \hfill
    \begin{subfigure}[b]{.48\linewidth}
        \centering
        \includegraphics[width=1.0\linewidth]{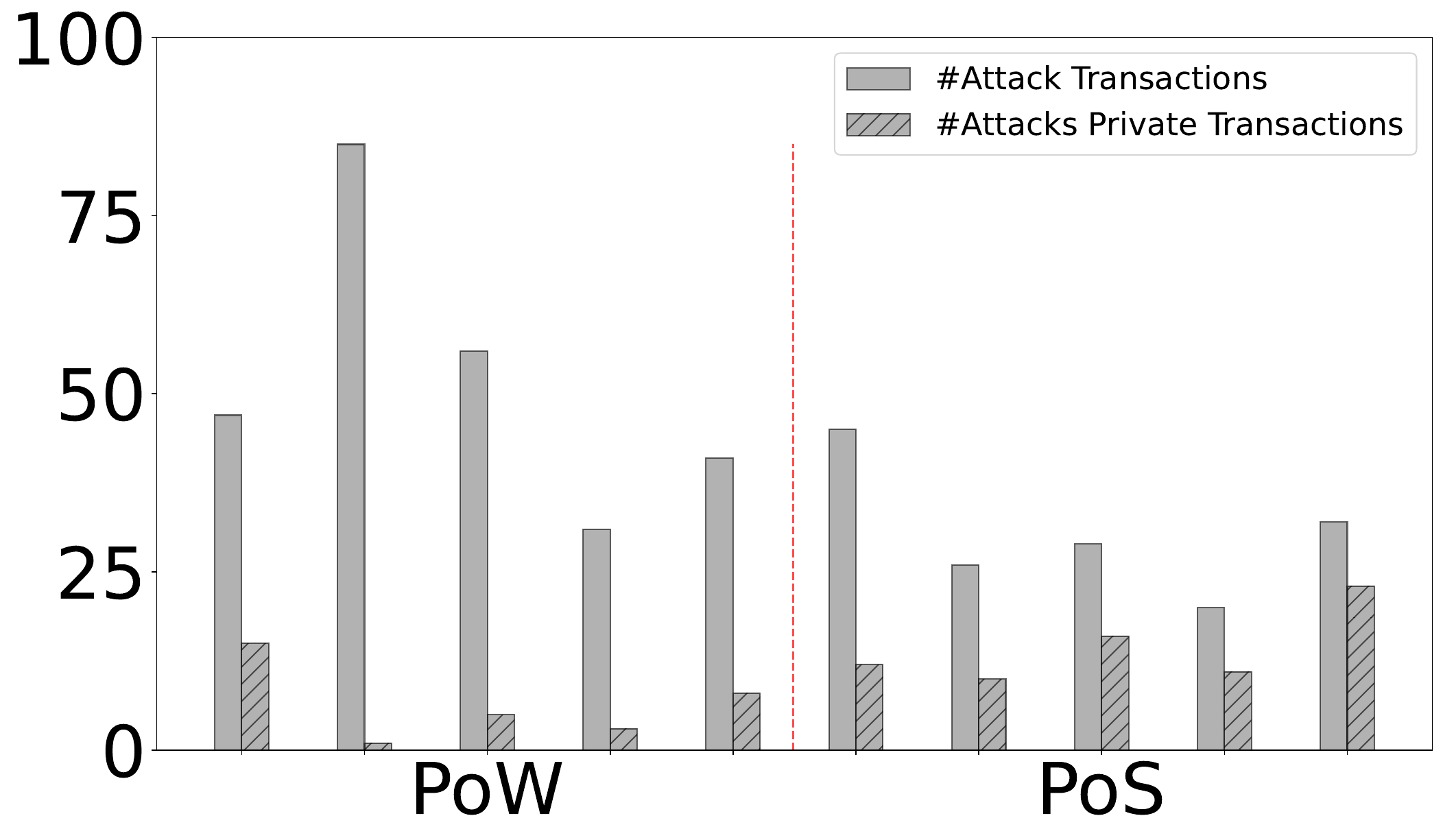}
        \caption{Count of attack transactions}
        \label{figs:defi_ptxs}
    \end{subfigure}
    \caption{Summary of DeFi attacks and private transactions.}
\end{figure}

\figref{figs:defi_attacks} illustrates the count of DeFi attacks and the count of attacks utilizing private transactions, aggregated over every 3-month period. Note that an additional half month is incorporated into the last 3-month period for both PoW and PoS datasets. These figures reveal a noticeable increase in the proportion of private transactions used in attacks from the PoW to the PoS period, indicating a significant surge in the utilization of private transactions within DeFi attacks. On average, there are 38 attacks in total and 4 attacks utilizing private transactions per three months in PoW Ethereum, while in PoS Ethereum, there are 17 attacks and 7 attacks utilizing private transactions per three months. Moreover, one attack may involve multiple transactions. Therefore, we further explore the attack related transactions and the usage of private transactions in \figref{figs:defi_ptxs}.
In the PoW era, there are a total of 20 DeFi attacks involving 32 private transactions, whereas in the PoS era, 35 DeFi attacks utilize 72 private transactions.

Additionally, attacks in the PoW era yield an average profit of \$43.36 million per attack, while in the PoS era, the average earnings per attack decrease to \$8.64 million. Regarding profit distribution as bribes from attackers, miners received approximately 0.2541 \textit{ETH} per attack in PoW, which decrease to 0.0845 \textit{ETH} in PoS. Specifically, the most frequently targeted platforms are decentralized exchanges (DEX) with 28 reported incidents, followed by attacks on lending services (9), tokens (8), NFTs (5), bridges (4), and Gaming \& Metaverse (1).
% \mz{@xingyu, please redraw the figure and take a look at the description}

\vspace{-8pt}
\section{Transaction Economics}
\label{sec:eco}
We study the monetary flow of \pri \txs from users to blcok creators, especially measuring \tx cost spent by users and mining profits earned by block creators. We also evaluate the different entities participating in the mining process and the mining profits distribution. 

\subsection{Transaction Cost}

\subsubsection{Used gas} \figref{figs:usedgas} presents the used gas for \pri \txs per block in PoW and PoS Ethereum. 
We observe that the expansion of the gas limit after EIP-1559~\cite{eip1559} has a significant impact on private transactions, leading to an increase in their number and relieving network congestion. We also observe that the average used gas per private transaction in PoS increases by 26.06\%, compared to that of PoW Ethereum. In PoW Ethereum, the average used gas per private transaction is 125,658 units, with 735 blocks reaching 100\% gas usage. Out of these 735 blocks, 629 were mined after the expansion of the gas limit, while the remaining 106 were mined before. Conversely, in PoS Ethereum, the average used gas per private transaction increases to 159,427 units and there are 253 blocks with full gas usage (after the expansion).

\subsubsection{Gas price} 
\begin{figure}[t]
    \centering
    \begin{subfigure}[b]{.47\linewidth}
        \centering
        \includegraphics[width=\linewidth]{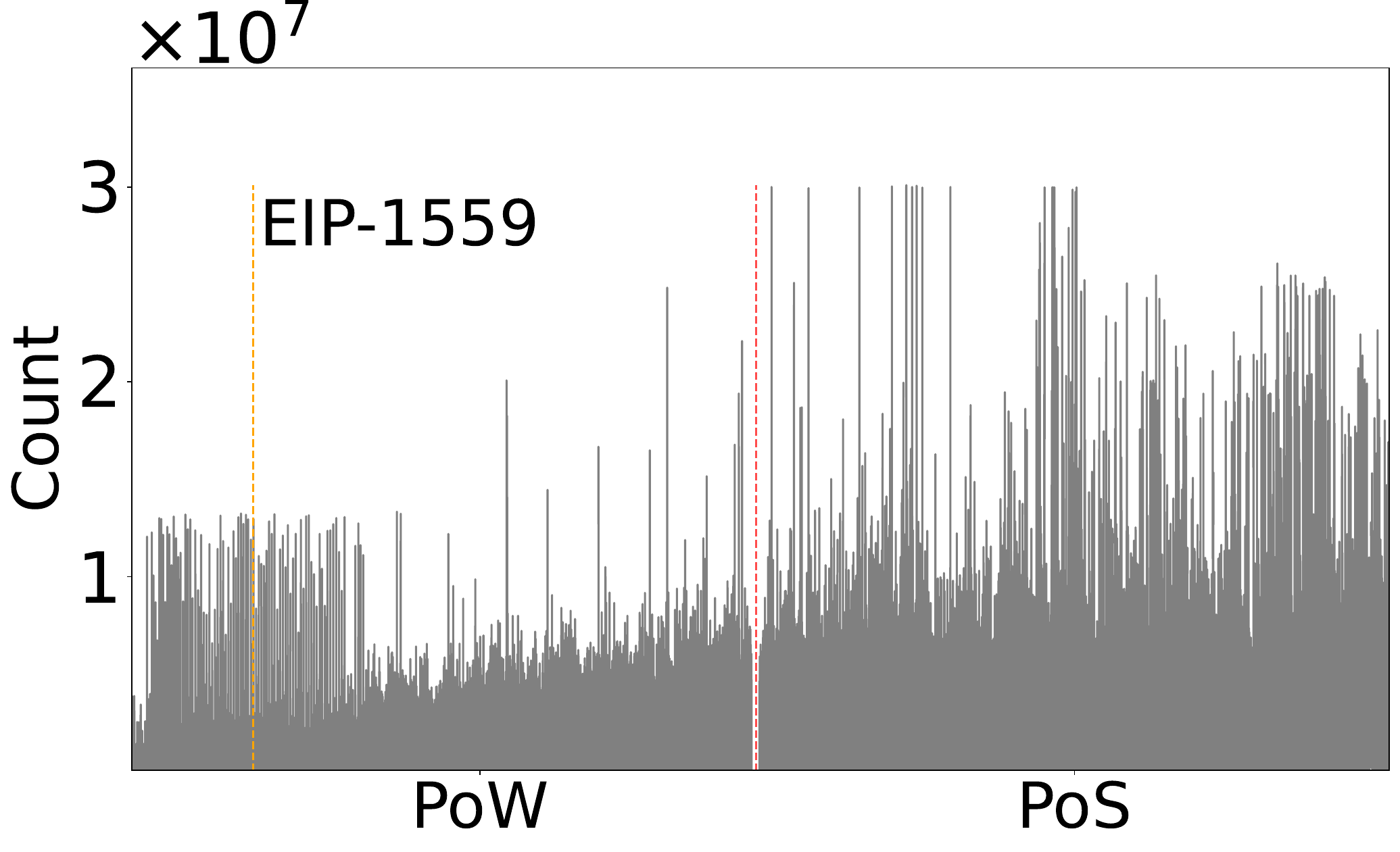}
        \caption{Used gas}
        \label{figs:usedgas}
    \end{subfigure}
    \hfill
    \begin{subfigure}[b]{.48\linewidth}
        \centering
        \includegraphics[width=\linewidth]{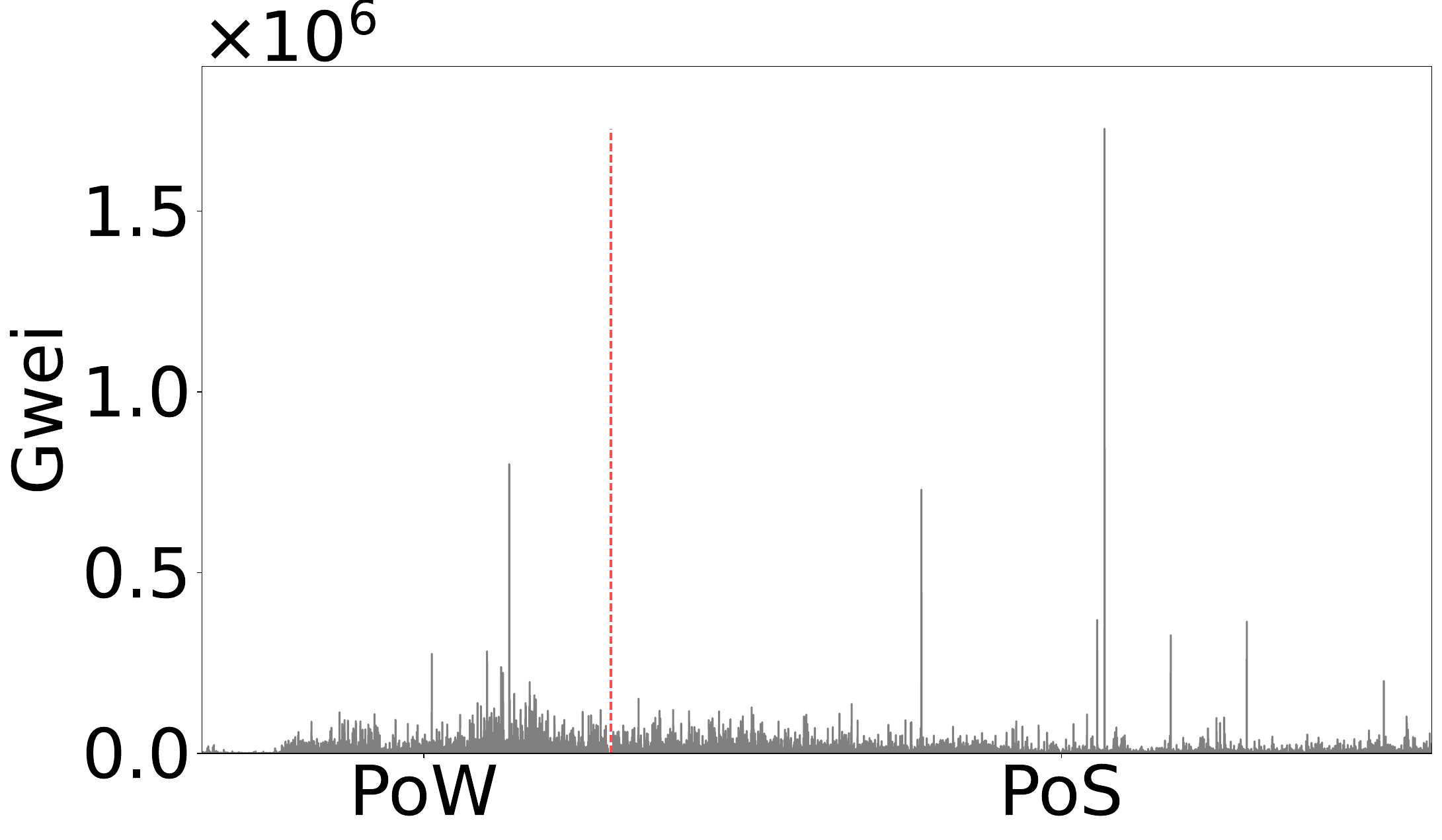}
        \caption{Gas price}
        \label{figs:gasprice}
    \end{subfigure}
    \caption{Transaction fee per \pri \tx.} 
\end{figure}

\figref{figs:gasprice} presents the gas price for each private transaction in PoW and PoS Ethereum.
We observe that \pri \txs in PoS Ethereum have much lower gas price, compared to both pre and post EIP-1559 gas prices in PoW Ethereum. Moreover, \pri \txs with 0 priority fee in both PoW and PoS Ethereum mainly call MEV Bots to search for MEV opportunities.  In PoW Ethereum, the implementation of EIP-1559 has a significant impact on the gas price. Prior to EIP-1559, there are 3,507,467 private transactions, with an average gas price of 97.3 \textit{Gwei}. Approximately 50\% of these transactions have 0 gas price. After EIP-1559, there are 11,302,925 \pri \txs and their average gas price increases to 203.0 \textit{Gwei}. With required basefee, there are no more \pri \txs with 0 gas price, but 26\% have 0 priority fee. In PoS Ethereum, the average gas price for private transactions is significantly lower at 60.98 \textit{Gwei}. Additionally, there are 16.19\% private transactions with 0 priority fee, with a majority of them calling MEV Bots.

\subsubsection{Transaction fee} 
\begin{figure}[t]
    \begin{subfigure}[b]{.48\linewidth}
        \centering
        \includegraphics[width=\linewidth]{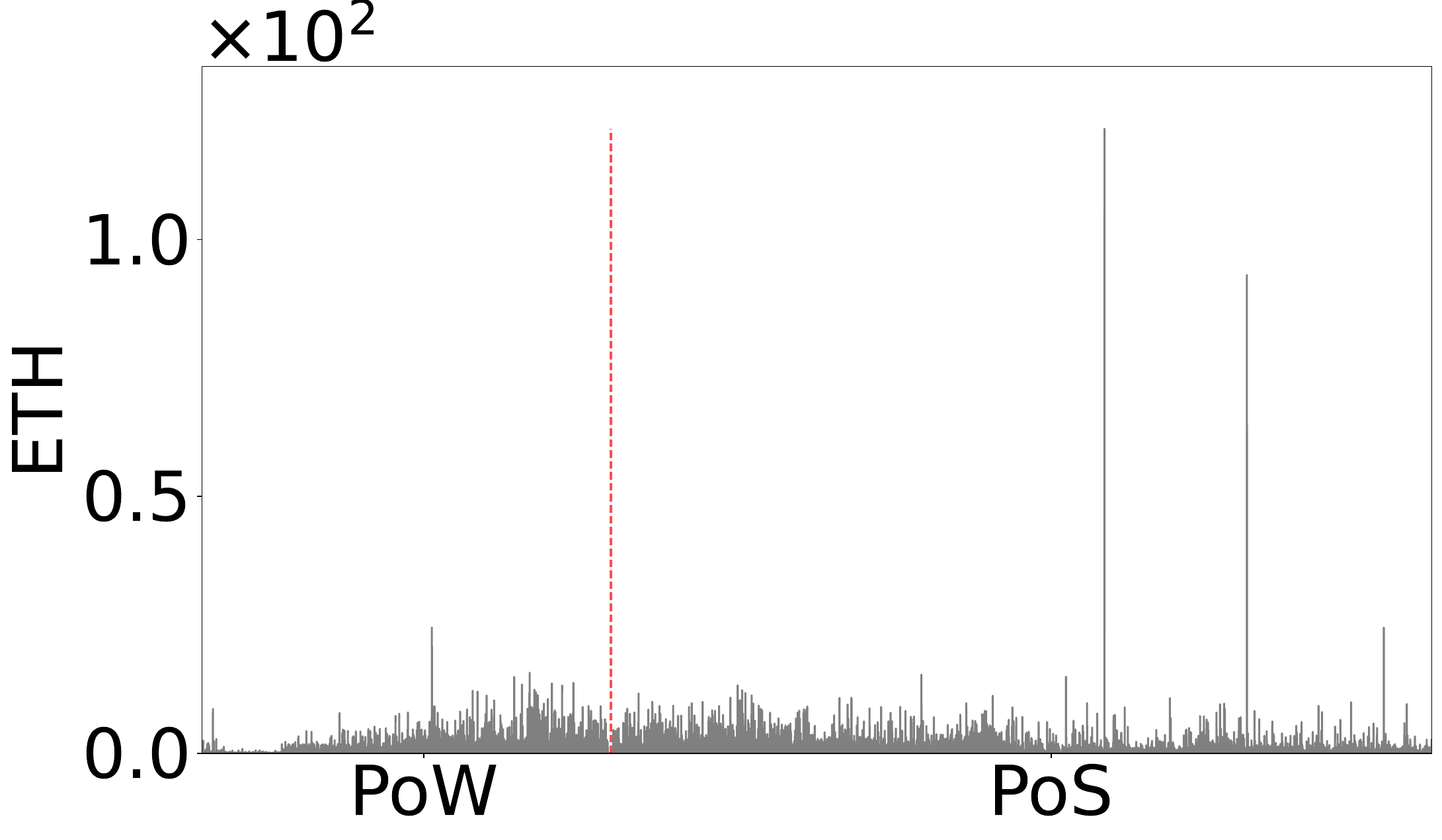}
        \caption{\Tx fee}
        \label{figs:pow_fee_per_px}
    \end{subfigure}
    \hfill
    \begin{subfigure}[b]{.46\linewidth}
        \centering
        \includegraphics[width=\linewidth]{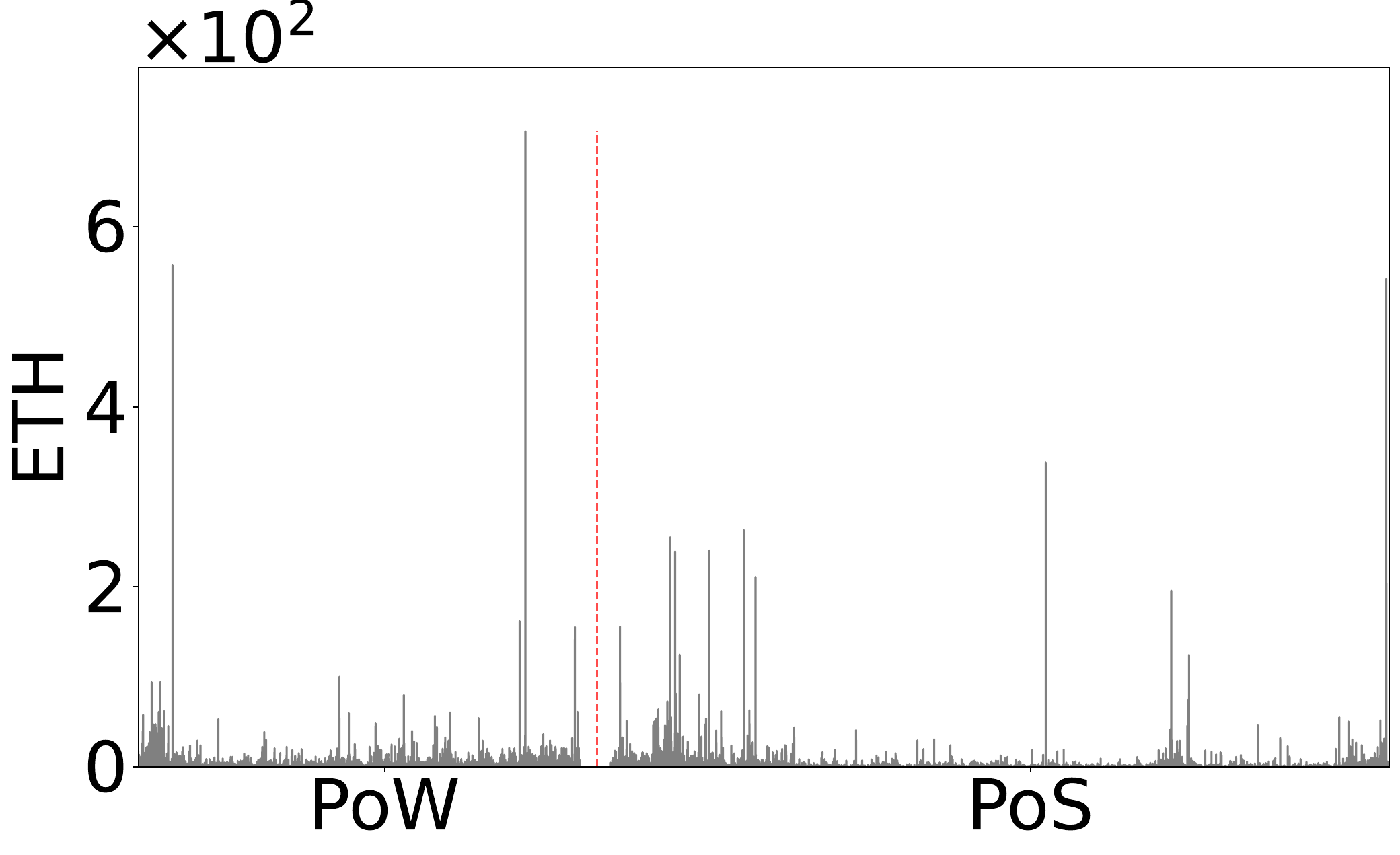}
        \caption{Direct payment}
        \label{figs:pow_payment_per_px}
    \end{subfigure}
    \caption{Transaction cost of \pri \txs per block.}
\end{figure}

\figref{figs:pow_fee_per_px} displays the transaction fee for each private transaction in PoW and PoS Ethereum.
We observe that the \tx fee of \pri \txs in PoS Ethereum is almost half of that of PoW Ethereum on average, mainly due to the lower gas price. We also observe there are more extremely large \tx fee peeks in PoS Ethereum, which might be caused by the high competition between \tx users. In PoW Ethereum, the average \tx fee of all the \pri \txs is 0.0183 \textit{ETH}. The largest transaction fee, 24.50 \textit{ETH}, is from transaction~\texttt{0xf769aacc}, which exchanges 34,254.56 \textit{RGT} tokens (\$13,523.17) for 201.57 \textit{ETH} (\$695,271.24) for a profit. In PoS Ethereum, the average \tx fee is 0.0105 \textit{ETH}. We also find three \pri \txs have at least 24.50 \textit{ETH}. The top 1 private transaction~\texttt{0x4a84d087} has 121.56 \textit{ETH} \tx fee.

% We evaluate the transaction fee, considering both the used gas and gas price, as \textit{TxFee = UsedGas × GasPrice}.

% \begin{figure}[t]
%     \centering
%     \begin{subfigure}[b]{.47\linewidth}
%         \centering
%         \includegraphics[width=\linewidth]{figs/raid/gas_used_per_blocks.pdf}
%         \caption{Used gas}
%         \label{figs:usedgas}
%     \end{subfigure}
%     \hfill
%     \begin{subfigure}[b]{.48\linewidth}
%         \centering
%         \includegraphics[width=\linewidth]{figs/raid/price_total.pdf}
%         \caption{Gas price}
%         \label{figs:gasprice}
%     \end{subfigure}
%     \caption{Transaction fee per \pri \tx.} 
% \end{figure}

\subsubsection{Direct payment} \figref{figs:pow_payment_per_px} shows the direct payment for each private transaction in PoW and PoS Ethereum. 

We observe that the average direct payment decreases by 85.71\%, while the percentage of \pri \txs making direct payments increases by 10.97\% in PoS Ethereum, compared to that of PoW Ethereum. In PoW Ethereum, there are 1,540,000 (10.39\%) \pri \txs make direct payment, with an average value of 0.07 \textit{ETH}. The lowest direct payment is made by transaction~\texttt{0x35c510d6} (1.00e-9 \textit{ETH}) and the highest is made by transaction~\texttt{0x3b5fc9f8} (706.33 \textit{ETH}), which is also the highest peak in the figure. In PoS Ethereum, there are 1,403,814 (4.67\%) \pri \txs make direct payment, with an average value of 0.01 \textit{ETH}. 

% \lesson{PoS Ethereum has lower private transaction cost compared to PoW Ethereum, with average decreases of 42.62\% in transaction fee and 85.71\% in direct payment. 
% } 

\begin{figure*}[t]
    \begin{subfigure}[b]{.32\linewidth}
        \centering
        \includegraphics[width=\linewidth]{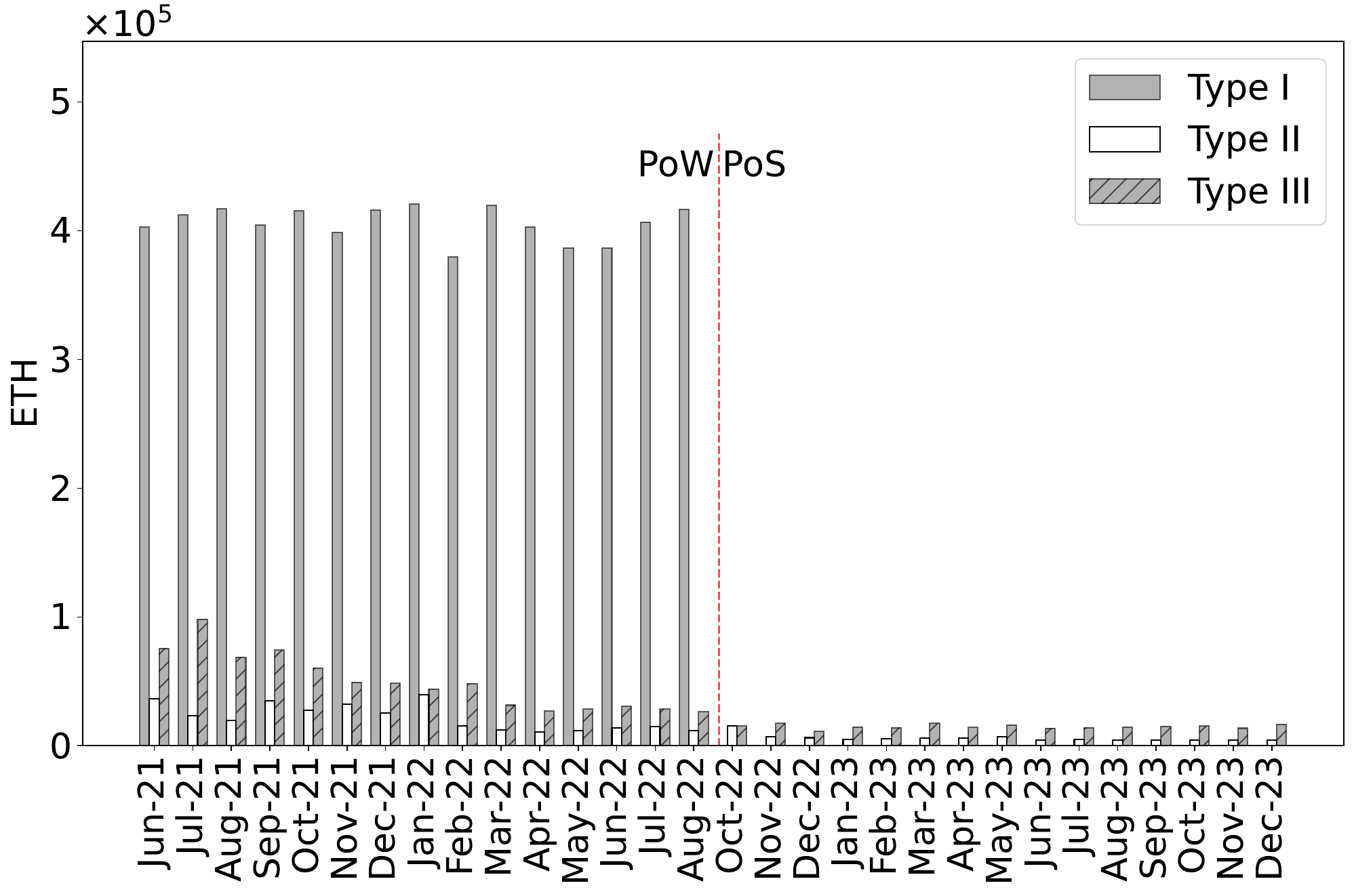}
        \caption{Three types}
        \label{figs:mt}
    \end{subfigure}
    \hfill
    \begin{subfigure}[b]{.32\linewidth}
        \centering
        \includegraphics[width=\linewidth]{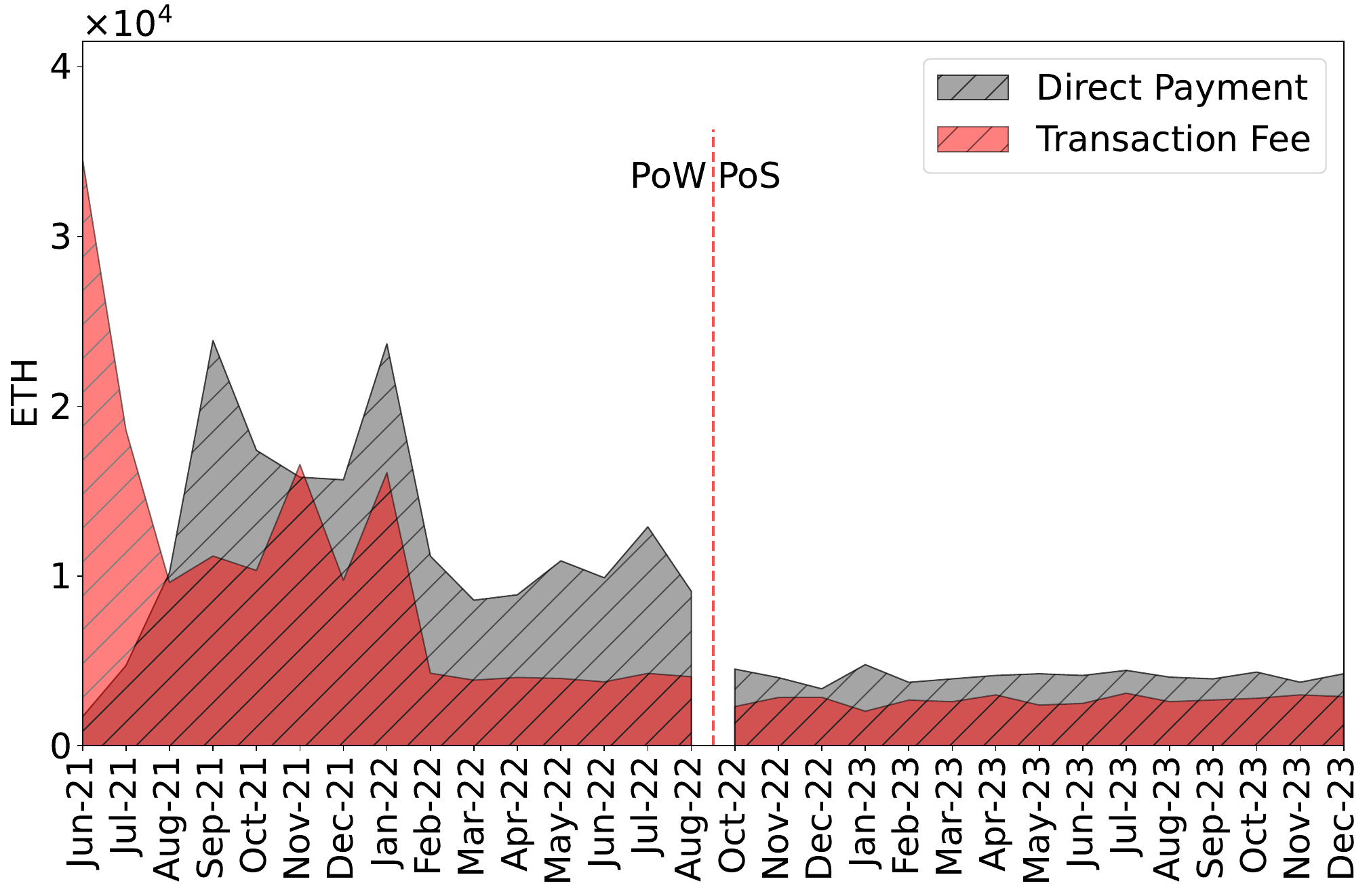}
        \caption{Type II (private transaction)}
        \label{figs:ptp}
    \end{subfigure}
    \hfill
    \begin{subfigure}[b]{.32\linewidth}
        \centering
        \includegraphics[width=\linewidth]{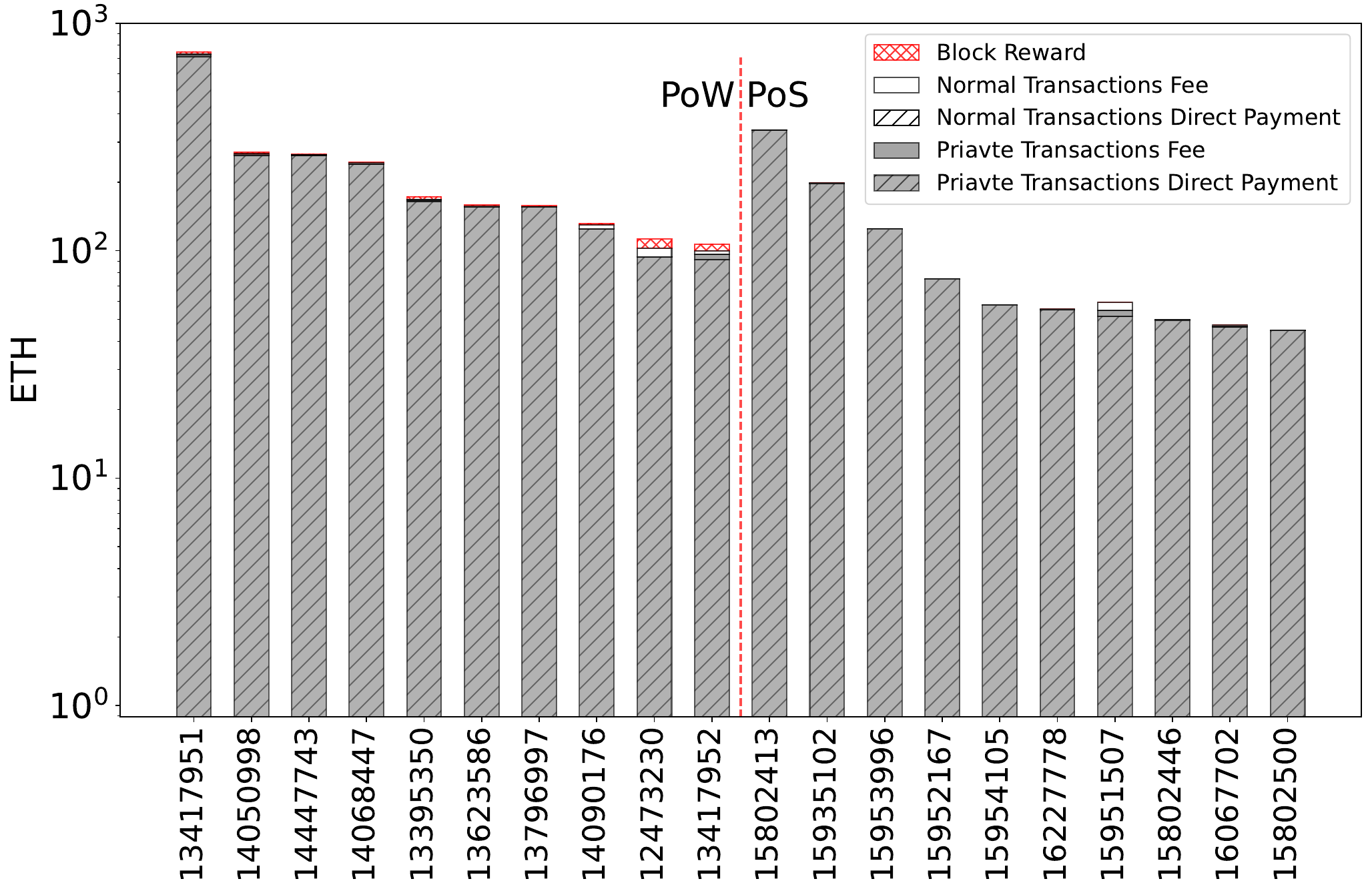}
        \caption{Top 10 blocks}
        \label{figs:top_blocks}
    \end{subfigure}
    \caption{Mining profits. 
    }
    \vspace{-10pt}
\end{figure*}

\subsection{Mining Profits}

\subsubsection{Profits Classification}
To evaluate the influences of private transactions,
we classify mining profits into three types.
\begin{packeditemize}
    \item Type I: Block rewards. The rewards consist of both normal and uncle blocks payouts.
    \item Type II: Profits from \pri \txs. The profits include both transaction fee and the direct payment in \pri \txs. Specifically, basefee in transaction fee will be burnt by the network and deducted from the mining profits.
    \item Type III: Profits from normal \txs (similar to Type II). 
\end{packeditemize}

\subsubsection{Three Types Profits} 
\figref{figs:mt} presents the monthly mining profits categorized into three types. 

We observe a significant decline in mining profits during the PoS era. This decline can be attributed primarily to the absence of Type I rewards in PoS Ethereum, historically the most substantial contributor to mining profits (which amounted to 421,438 \textit{ETH} per month in PoW Ethereum). Additionally, profits from both Type II (6,014 \textit{ETH} per month) and Type III (7,486 \textit{ETH} per month) are diminished in the PoS era when compared to the figures of 30,510 \textit{ETH} per month for Type II profits and 99,968 \textit{ETH} per month for Type III profits in the PoW era. 
We speculate that the transition to the new PoS consensus mechanism has alleviated network congestion, resulting in reduced tips for block creators. Additionally, the emergence of private transaction service providers has led to a decrease in direct payments to block creators. Another potential factor is the support for offline payments by some private transaction service providers, allowing users to directly compensate validators without incurring transaction fees (e.g., BloXroute~\cite{blockxroute_offline}).

\subsubsection{Type II Profits (Private Transaction)}
\figref{figs:ptp} specifically illustrates the monthly profits from Type II, which encompasses transaction fees and direct payments from private transactions.

We observe that, in most cases, profits derived from the direct payment method in private transactions surpass those from transaction fees, both in the PoW and PoS eras. This trend is particularly pronounced in the PoS era, where direct payments consistently yield higher profits. Moreover, we find that, in June 2021 (the first month of our PoW dataset), there is a spike in transaction fee profits due to increased private transaction usage.

\subsubsection{Top 10 block profits} 
\figref{figs:top_blocks} displays the top 10 blocks that generate the highest profits.

We observe that the direct payments from \pri \txs take a majority of profits for every top 10 block and other profits contribute a rather small percentage, in both PoW and PoS Ethereum. 
In PoW era, the top 1 block 13,417,951 earns around 747 \textit{ETH}, with 714 \textit{ETH} coming from private transactions. The most profitable private transaction in this block transferred 706.33 \textit{ETH} directly to the miner. In PoS era, the top 1 block is 15,802,413, which earns 337.96 \textit{ETH} in total and the direct payment from one \pri \tx accounts for 99.99\% (337.92 \textit{ETH}).

We also observe that the private transaction sending the most direct payment within the top 10 blocks is either MEV related or DeFi attack related. In PoW era, all the most profitable private transactions are MEV transactions, namely arbitrage and liquidation. In PoS era, such \pri \txs are all related to MEV and attacks. For example, The attack private \tx in block 16,227,778 caused a loss of \$82,393.42 at that time.

% \lesson{Private transactions can yield substantial profits for block creators in both the PoW and PoS eras, such as million-dollar profits from one private transaction. This is noteworthy, especially considering the significantly lower average mining profits in the PoS era, primarily due to the elimination of block rewards.} 

\subsection{Mining Participation}
\label{app:eco:par}
We assess the involvement of various participants (block builders, relays, and miners/validators), in the mining process of private transactions. It is important to note that in PoS, mining profits may not be directly received by validators. Instead, these profits are initially transferred to the block builders, who subsequently distribute the collected fees to the validators.

\subsubsection{Block Builders}
% Miners create new blocks in PoW era, while mostly, block builders construct new blocks and validators propose such blocks in PoS era. 
\tabref{tab:top_builders} lists top 10 block builders, ranked by the number of blocks with at least one private transaction they have created in PoS Eethereum.

\begin{table}[t]
    \begin{center}
    \setlength{\tabcolsep}{2pt}
    \resizebox{1\linewidth}{!}{
    \footnotesize
    \begin{tabular}{ rrrr } 
    \hline
    \scriptsize \textbf{Block Builder} & \scriptsize \textbf{\makecell{Address}} & \scriptsize \textbf{ \makecell{ \#Total Blocks}}  & \scriptsize \textbf{\makecell{ \#Blocks w. Private Tx}} \\ 
    \hline
    \scriptsize  Flashbots:Builder & \scriptsize 0xDAFEA4   & \scriptsize 280,251  & \scriptsize 280,251 (100.00\%)  \\ 
    \scriptsize  Builder0x69 & \scriptsize 0x690B9A   &  \scriptsize 240,028 &  \scriptsize 202,459 (84.35\%) \\
     \scriptsize  Beaverbuild  &  \scriptsize 0x952222   &   \scriptsize 190,637  & \scriptsize 102,897 (53.98\%) \\
     \scriptsize  Lido: Exec Layer Rewards Vault  &   \scriptsize 0x388C81   & \scriptsize 59,631  & \scriptsize 12,305 (20.64\%) \\
     \scriptsize  BloXroute:Max Profit Builder  & \scriptsize 0xF2f5C7   & \scriptsize 46,797  &  \scriptsize 46,797 (100.00\%) \\

    \scriptsize  Fee Recipient:0x467...263   & \scriptsize 0x4675C7   &  \scriptsize  44,639  & \scriptsize 8,881 (19.90\%) \\

    \scriptsize  Eth-builder  &  \scriptsize 0xFeebab &  \scriptsize 36,437  & \scriptsize 18,408 (50.52\%) \\
     \scriptsize  MEV Builder: 0x473...dFc   & \scriptsize 0x473780   &  \scriptsize 29,456  & \scriptsize 24,247 (82.32\%) \\

    \scriptsize  BloXroute:Regulated Builder  & \scriptsize 0x199D5E   &  \scriptsize  28,691  & \scriptsize 28,691 (100.00\%) \\
     \scriptsize  Eden Network:Builder  &  \scriptsize 0xAAB27b   & \scriptsize 23,684  &  \scriptsize 23,684 (100.00\%) \\
    \hline
    \end{tabular}}
    \caption{Top 10 block builders in PoS Ethereum.
    }
    \label{tab:top_builders}
    \vspace{-10pt}
    \end{center}
\end{table} 

We observe that 4 of top 10 block builders only mine blocks with \pri \txs, including  \textit{Flashbots:Builder}, \textit{BloXroute:Max Profit Builder}, \textit{BloXroute:Regulated Builder}, and \textit{Eden Network:Builder}. Moreover, we observe that some block builders mainly generate blocks without \pri \txs, including \textit{Lido:Execution Layer Rewards Vault} and \textit{Fee Recipient:0x467}, which only mine 20.64\% and 19.90\% blocks with \pri \txs, respectively. We also observe that the top 3 block builders produce far more blocks than other builders, such as the number of mined blocks top 3rd produce is almost 3 times of that of the top 4th block builder.

Moreover, we observe that certain block builders are affiliated with the same private transaction service provider, while others have no specific connections. For instance, the second builder, \textit{BloXroute:Max Profit Builder}, and the ninth builder, \textit{BloXroute:Regulated Builder}, are exclusively associated with BloXroute. In contrast, \textit{Builder0x69} has the flexibility to connect with any private transaction service provider, such as Flashbots or Ultra Sound~\cite{mevboost}.

\subsubsection{Relays} A relay serves as a non-trivial component in \pri \tx service provider, in both PoW and PoS Ethereum. We observe that although the \textit{Flashbots Relay} is still the top 1 relay used to mine \pri \txs, it takes a much smaller percentage (35.64\%) in PoS era compared to the percentage (82.9\%) in PoW era, due to the emerging of \pri \tx service providers in PoS era. In PoW era, Flashbots is the leading private transaction service provider connecting most miners~\cite{flashbots-pow}. To study the impacts of Flashbots on \pri \txs, we collect Flashbots blocks by querying their APIs~\cite{flashbots-api} and obtain 1,043,919 blocks, which accounts for 82.9\% among all the blocks. In PoS era, we fetch the block number and its related relay address from beaconcha~\cite{relays} to study relays. The top 10 relays are Flashbots Relay (35.64\%), Ultra Sound(23.82\%), BloXroute Max-Profit Relay(16.52\%), Agnostic Gnosis(10.37\%), Blocknative Relay(4.36\%), BloXroute Regulated Relay (2.53\%), Eden Network (2.12\%), BloXroute Ethical Relay(0.66\%), Aestus(0.56\%), ManiFold(0.34\%), respectively. 

Similar to block builders, we also observe some relays belong to the same party. For example, BloXroute owns three MEV relays: \textit{BloXroute: Max Profit relay}, \textit{BloXroute: Regulated relay}, and \textit{BloXroute: Ethical relay}. Our investigation uncovers the presence of 14 relays in the ecosystem, with Flashbot emerging as the foremost and most established participant, commanding a significant market share. Conversely, Ultra Sound~\cite{ultra-sound} is a comparatively recent addition, with available data spanning from December 2022 onwards, yet it demonstrates promising growth prospects. Notably, we observe that a single relay has the capability to connect with multiple builders during the mining of a new block. For instance, Flashbots accommodates transactions submitted by more than 20 distinct block builders~\cite{relayscan}.

\subsubsection{Miners/Validators}
% In~\tabref{tab:top_miners}, w
We study the top 10 miners and validators in PoW and PoS Ethereum, respectively, sorted by the number of their mined blocks containing \pri \txs.

We observe that \textit{Ethermine} dominates the miner market in PoW era and \textit{Lido} dominates the market in PoS era. In PoW Ethereum, we identify 435 unique miner address and the 1st miner \textit{Ethermine} mines 316,774 blocks with \pri \txs, which is almost two times of the 2nd miner \textit{F2Pool}. In PoS Ethereum, we identify 61,148 unique validators. \textit{Lido} mines the most block (170,525), which is almost three times of the 2nd validator \textit{Fee Recipient: 0x467}. 

We also obverse that miners in PoW era usually send lots of \pri \txs for redistributing mining profits, while validators in PoS era usually receive many \pri \txs for obtaining mining profits from block builders. For example, \textit{Lido} receives the same amount (170,525) of \txs as the number of mined blocks with \pri \txs. We suspect that all the blocks with \pri \txs mined by \textit{Lido} go through some \pri \tx service providers. Thus, every block builder sends one \pri \tx to \textit{Lido} for every mined block.

% \lesson{\Pri \txs are less centralized in PoS era, 
% % in terms of participation of participants in the mining process, 
% due to the participation of many new \pri \tx service providers recently. Flashbots serves as the leading \pri \tx service provider with 82.9\% market share in PoW era; whereas in PoS era, Flashbots only takes 35.6\% market share.}

\subsection{Profits Distribution}
\label{app:eco:dis}
We perform a case study to figure out the miner/validator profit distribution in PoW and PoS Ethereum, respectively.

We observe that miners in the PoW era frequently transfer \textit{ETH} to the mining nodes as rewards for their computational power. These profit redistribution transactions are sent as private transactions because miners can avoid long waiting times and save money on priority fees by sending transactions to themselves.

To examine the flow of mining profits, we perform a case study of the top 1 miner, \textit{Ethermine}, with 587,656 (3.97\%) private transactions. 
% \tabref{tab:miner_flow} lists
We study the top five addresses receiving
the most private transactions from Ethermine, which are all
EOAs. Most of these EOAs act as wallets for miner nodes, receive profits from \textit{Ethermine}, and then withdraw money. For instance, address~\texttt{0x909bfe97} receives 509 transactions, including 58 private transactions, and sends 152 transactions to withdraw money from its personal wallet through DeFi applications. The behavior is similar for the other four EOAs.

% \begin{table}[t]
%     \centering
%     \resizebox{1\linewidth}{!}{
%     \footnotesize
%     % \setlength{\tabcolsep}{4pt}
%     \begin{tabular}{lccccc}
%       \toprule
%       \multirow{2}{*}{Address} & \multicolumn{3}{c}{\Tx Inbound} & 
%       \multicolumn{2}{c}{\Tx Outbound} \\ 
%       \cline{2-4} \cline{5-6}
%       & \#Total & \#Ethermine & \#Private & \#Total & \#Exchange \\
%       \midrule
%       \scriptsize  0x909bFe97~\cite{0x909bfe97}  & \scriptsize 509 &\scriptsize  509 &\scriptsize 58 &\scriptsize  152 &\scriptsize  152 \\
%    \scriptsize  0x18690F0E~\cite{0x18690F0E} & \scriptsize 492 & \scriptsize 468 &\scriptsize 67 &\scriptsize  191 & \scriptsize 185  \\
%    \scriptsize  0xE426ec22~\cite{0xE426ec22} &\scriptsize  476 & \scriptsize 476 &\scriptsize 64 &\scriptsize  49 &\scriptsize  49  \\
%    \scriptsize  0xa4C916F3~\cite{0xa4C916F3} &\scriptsize 327 &\scriptsize  312 &\scriptsize 72 & \scriptsize 23 & \scriptsize 23  \\
%    \scriptsize  0xbaAa2853~\cite{0xbaAa2853} &\scriptsize  341 &\scriptsize  341 &\scriptsize 52 &\scriptsize  4 &\scriptsize  0 \\
%       \bottomrule
%     \end{tabular}}
%     \caption{Top 5 addresses receiving the most \pri \txs from the miner Ethermine in PoW Ethereum. 
%     \textit{\Tx Inbound:} the \txs sent to the address. 
%     \textit{\Tx Outbound:} the \txs sent from the address. 
%     \textit{\#Exchange:} the count of \txs that are used for exchanging tokens.} 
%     \label{tab:miner_flow}
% \end{table}

We observe that validators in the PoS era may accumulate mining profits for a period of time before transferring them to other accounts. For example, the validator~\texttt{0x4675C7e5} sent a cumulative amount of \textit{ETH} to Coinbase 10~\texttt{0xA9D1e08C}, after receiving mining profits from block builders. We analyze the \txs related to Coinbase 10 and observe that, block builders typically use \pri \txs to transfer money to validators; whereas, validators use normal \txs to send money to others.

% \lesson{In PoW Ethereum, miners directly obtain mining profits from users and then distribute the profits among participants according to their contributions of mining power cost; whereas in PoS Ethereum, validators usually receive profits from block builders and then collect the left fees to their belonging parties (e.g., Coinbase).}

\section{Discussion}
\label{sec:dis}

\bheading{Private transaction datasets.}
We source private transactions from Blocknative~\cite{Blocknative}, a platform also utilized in prior studies~\cite{adams2023costs, heimbach2023ethereum}. Blocknative employs a network of Ethereum nodes worldwide to detect private transactions; if a transaction is not observed by any node but is still confirmed in a block, it is classified as a private transaction. This approach has also been employed by \textit{Sen} et al.~\cite{yang2022sok}. Due to resource constraints for extensive, multi-node monitoring over extended periods, we chose to utilize the dataset provided by Blocknative. To validate the accuracy of the Blocknative private transaction datasets, we manually verified a subset of 100 randomly selected private transactions, all of which were found to be 100\% accurate. However, occasional misclassifications of private transactions due to network latency may occur, which we acknowledge as a limitation. 
% To ensure accuracy, Blocknative operates multiple global nodes for transaction monitoring.

\bheading{Private \txs on other blockchains.}
Similar as Ethereum, Binance Smart Chain (BSC) is built based on Ethereum Virtual Machine (EVM) and smart contracts. However, BSC uses a consensus of Proof of Staked Authority (PoSA)~\cite{bsc} and there are only 21 validators. 
Recently, BNB48 has provided a \pri \tx service (Enhanced RPC) on BSC~\cite{bnb48-ptx}, but is not as widely used as in Ethereum blockchain.
For other blockchains, such as Polygon, which can be considered as a fork of Ethereum, uses a Proof-of-Stake (PoS) consensus mechanism.
To the best of our knowledge, there is no official \pri \tx concept in the Polygon, while there are MEV Bots. 
We believe the reason is that there are no special channels in Polygon (e.g., FlashBots RPC in Ethereum) used for \pri \txs. 
We believe that there is more to observe and examine in such blockchains about MEV and \pri \txs.

\section{Related Work}
\label{sec:related}
We examine related works that focus on private transactions, Ethereum transactions, Ethereum interaction networks.

\bheading{Private Transactions.} 
Private transactions remain relatively understudied in the research community. Only a few works~\cite{piet2022extracting,weintraub2022flash,capponi2022evolution,lyu2022empiricalstudyethereumprivate} have addressed private transactions, primarily as by-products while investigating MEV in PoW Ethereum. Additionally, \textit{Yang et al.}~\cite{yang2022sok} utilize private transactions to assess 28 MEV auction platforms in PoS Ethereum. In contrast to these works, our research focuses primarily on private transactions, while they concentrate on MEVs. Furthermore, most of these works exclusively examine PoW Ethereum, with only a few exploring PoS Ethereum, whereas our study encompasses both PoW and PoS Ethereum to provide comprehensive comparative insights.

\textit{Piet et al.}~\cite{piet2022extracting} measured \pri \txs in transaction pools from four sources for about 1 day, and found that most of the \pri \txs are used for miner profit redistribution and MEV extraction. In particular, 91.5\%  of MEV extraction is done via \pri \txs. They also analyzed the miner profits among \pri \txs and the top five miners earned most profits. 
Besides, 40\% of miners had not mined any \pri \tx. 
We are different in the focuses of analysis.
For example, they measure the inconsistency of \pri \txs in miners, while our work focuses on the profits flows earned by miners. 
Besides, we have a much larger dataset (one year) and we present a study on the security impacts of private \txs.  

\textit{Weintraub et al.}~\cite{weintraub2022flash} mainly measure and analyze the impact of Flashbots with focusing on MEV, which provides a private channel between Ethereum users and miners. \textit{Weintraub et al.} also measures other private services providers (e.g., Eden Network~\cite{edennetwork}) and shows that most of the MEV extraction comes from private service providers, especially Flashbots. They discover the unfairness of Flashbots, in which miners have profited much more than MEV searchers.
Their work studies \pri \txs mainly from the perspective of MEV extraction, while our work gives a more complete view on \pri \txs in other aspects.

\textit{Qin et al.}~\cite{qin2021quantifying} measure Blockchain Extraction Value (BEV) to detect sandwich attacks, liquidation, and arbitrage in \txs. Specifically, they consider the \tx with zero gas price as \pri \txs, and measure the percentage and value in each BEV category. Moreover, they formalize BEV relay systems and analyze the impacts on P2P network and consensus layer. They mainly study BEV, while our work highlights \pri \txs. 

\textit{Capponi et al.}~\cite{capponi2022evolution} propose  a game theoretical model to analyze the economic incentives behind the venues from the \pri \txs. In particular, their paper collected and investigated the transactions summited via the private channel from the Flashbots. The results show that the implications on different entities, including the increasement of users and miners payoffs. Their work highlights the economic incentives of \pri \txs, while our work focuses on analyzing the behavior of \pri \txs and their impacts.

\bheading{Ethereum \txs.} 
\cite{said-tx,lin-tx,zanelatto2020transaction} study \txs via graph analysis to investigate the \txs behaviors. 
Specifically, \textit{Said et al.}~\cite{said-tx} analyze transaction behaviors, community structure and link prediction. ~\textit{Lin et al.}~\cite{lin-tx} model \tx records as temporal weighted multi-digraphs. \textit{Zanelatto et al.}~\cite{zanelatto2020transaction} conduct a study on 3-year Ethereum \txs with crypto currency. 
These works measure the overall features of \txs, while we focus on measuring \pri \txs.

\bheading{Ethereum Interaction Networks.}
To study the behaviors of different entities and their interactions, many works study Ethereum networks between users and smart contracts. Particularly, \textit{Lee et al.}~\cite{lee-measure}, ~\textit{Chen et al.}~\cite{chen2020understanding}, ~\textit{Lin et al.}~\cite{Zhao2021TemporalAO}, and ~\textit{Bai et al.}~\cite{bai-temporal} perform a measurement study on these interaction networks to observe the activities of Ethereum and propose some insights,
All of these works measure the general features of the Ethereum interaction networks between users and contracts from graph analysis, while our paper focuses on the purposes and usages of \pri \txs from the interaction networks.

% \bheading{EIP-1559.}
% \cite{liu2022empirical,Leonardos-eip1559,D-eip1559} conduct an empirical study on the effects of the EIP-1559 protocol. 
% \textit{Liu et al.}~\cite{liu2022empirical} lead an empirical study the effects of EIP-1559 protocol on transaction fee dynamics, transaction waiting time, and security exploits.
% \textit{Stefanos et al.}~\cite{Leonardos-eip1559} study both theoretical and experimental analysis of the dynamics and stability properties of the EIP-1559 protocol.
% \textit{Daniel et al.}~\cite{D-eip1559} study the fee market upgrade by leveraging the data from first month after the EIP-1559 protocol. They also propose an alternative rule to replace EIP-1559 and prove the effectiveness from the experimental results. 

\looseness=-1

% \section{Related Work}
% \label{sec:related}

% \bheading{Private Transactions.}
% Private \txs are still very new and have not received much attention in the research community. To date, there are only a few works~\cite{piet2022extracting,weintraub2022flash,capponi2022evolution,qin2021quantifying} that touch upon \pri \txs as by-products when investigating MEVs in PoW Ethereum.
% % (see~\appref{app:related:pri} for details). 
% Recently, \textit{Yang et al.}~\cite{yang2022sok} use \pri \txs to measure 28 MEV auction platforms in PoS Ethereum. Compared to these works, our work mainly studies the landscape of \pri \txs, while they focus on MEVs. Moreover, most of these works only focus on PoW Ethereum and a few study PoS Ethereum, while our work measures both PoW and PoS Ethereum for offering comparative insights.  

% \bheading{Ethereum \txs.} 
% Some works~\cite{said-tx,lin-tx,zanelatto2020transaction} study \txs via graph analysis to investigate the \txs behaviors. 
% Specifically, \textit{Said et al.}~\cite{said-tx} analyze transaction behaviors, community structure and link prediction. ~\textit{Lin et al.}~\cite{lin-tx} model \tx records as temporal weighted multi-digraphs. \textit{Zanelatto et al.}~\cite{zanelatto2020transaction} conduct a study on 3-year Ethereum \txs with crypto currency. 
% These works measure the overall features of \txs, while we focus on measuring \pri \txs.

% \looseness=-1

\section{Conclusion}
\label{sec:conclude}
We conduct the \textit{first} empirical comparative study on \pri \txs, to study their impacts in both PoW and PoS Ethereum.
To reveal the mysteries behind \pri \txs, we aim to answer three key questions regarding non-trivial aspects, including transaction characteristics, usages, and economic impacts. 
We obtain comparative findings from our measurement of \pri \txs in PoW and PoS eras, give suggestions to users based on our observations, and offer insights on the Ethereum community.
Overall, our work sheds light on the \pri \tx ecosystem and calls for attention on \pri \txs.

% \newpage
\bibliographystyle{ieeetr}
\bibliography{paper}

% % \clearpage
% \appendix
% \input{appendix-imc}

\end{document}